\pdfoutput=1
\documentclass[a4paper,11pt,english]{article}

\usepackage[bindingoffset=0.3cm,textheight=22.5cm,hdivide={2.7cm,*,2.7cm}, vdivide={*,22cm,*}]{geometry}
\usepackage{cite}
\usepackage{youngtab}
\usepackage{amsmath,amsfonts,amssymb,babel,slashed,color,empheq,tensor,amsthm}
\usepackage{hyperref}

\makeatletter

\newtheorem{thm}{Theorem}[section]

\newcommand{\del}{\partial}

\def\nn{\nonumber} 
\def\obar{\overline}
\numberwithin{equation}{section}
\def\a{\alpha}  \def\b{\beta}
 \def\g{\gamma} 
 \def\d{\delta} 

 \def\L{\Lambda}  
    \def\r{\rho}
  \def\t{\tau}

\def\cA{{\cal A}}  \def\cC{{\cal C}} 
\def\cD{{\cal D}}   
\def\cG{{\cal G}} \def\cH{{\cal H}} \def\cI{{\cal I}} 
 \def\cK{{\cal K}}  
\def\cM{{\cal M}} \def\cN{{\cal N}} \def\cO{{\cal O}} 
 \def\cQ{{\cal Q}}  
\def\cS{{\cal S}}

\def\R{{\mathbb R}} \def\C{{\mathbb C}} \def\N{{\mathbb N}}

 \def\one{\mbox{1 \kern-.59em {\rm l}}}

\def\msu{\mathfrak{su}}
\def\mso{\mathfrak{so}}

\newcommand{\Tr}{\mathrm{Tr}}

\newcommand{\End}{\mathrm{End}}

\def\Tr{\mbox{Tr}}

\newcommand{\eq}[1]{(\ref{#1})}

 \allowdisplaybreaks[3]

\begin{document}

%%%%%%%%%%%%%%%%%%%%%%%%%%%%%%%%%%%%%%%%%%%%%%%%%%%%%%%%%%%%%%%5%%%%%%%%%%%%%%%%%%%%%%%%%%

\parindent=0.4cm
\parskip=0.1cm

\renewcommand{\title}[1]{\vspace{10mm}\noindent{\Large{\bf#1}}\vspace{8mm}} 
\newcommand{\authors}[1]{\noindent{\large #1}\vspace{5mm}}
\newcommand{\address}[1]{{\itshape #1\vspace{2mm}}}

%%%% --- TITLE PAGE --- %%%%
\begin{titlepage}
\begin{flushright}
 UWThPh-2019-32
%  \\
%  \textcolor{red}{\today}
\end{flushright}
\begin{center}
\title{ {\Large On the quantum structure of space-time, gravity, and higher spin \\[1ex]
in matrix models}  }

\vskip 3mm

\authors{Harold C.\ Steinacker${}^\ddagger$}

\vskip 3mm

 \address{ 
${}^\ddagger${\it Faculty of Physics, University of Vienna\\
Boltzmanngasse 5, A-1090 Vienna, Austria  }  \\
Email: {\tt harold.steinacker@univie.ac.at}  
  }

\bigskip

\vskip 1.4cm

%%%% --- ABSTRACT --- %%%%
\textbf{Abstract}
\vskip 3mm

\begin{minipage}{14.5cm}%

In this introductory review, we argue that a quantum structure of space-time naturally entails a higher-spin theory, 
to avoid significant Lorentz violation. A suitable framework is provided by Yang-Mills matrix models,
which allow to consider space-time as a physical system, which is treated on the same footing as the fields that live on it. 
We discuss a specific quantum space-time solution, 
whose internal structure leads to a consistent and  ghost-free higher-spin gauge theory. 
The spin 2  modes give rise to metric perturbations,
which include the standard gravitons as well as the linearized Schwarzschild solution.

%  
% In this introductory review, we discuss a quantum space-time solution of Yang-Mills matrix models, whose
% internal structure leads to a consistent and  ghost-free
% higher-spin gauge theory. 
% The spin 2  modes give rise to metric perturbations,
% which include the standard gravitons as well as the linearized Schwarzschild solution.
%  The matrix model framework allows to consider 
% space-time as a physical system, which is treated on the same footing as the fields that live on it. 
%  

\end{minipage}

\end{center}

\end{titlepage}

\tableofcontents

\section{Introduction}

Our current understanding of fundamental physics is based on the concept of space-time,
which is assumed to be a pseudo-Riemannian manifold. 
This provides the stage for quantum field theory which describes all 
fundamental interactions except gravity, while 
gravity is described through the metric tensor on space-time which is governed by the 
Einstein equations.

However, there are 
good reasons to question the classical notion of space-time.
We know that quantum mechanics governs all physical matter and fields, and the quantum structure 
becomes more important at short distances. On the other hand, space-time 
and its metric is coupled to this quantum matter through the Einstein equations, and the coupling also becomes 
 stronger at short distances. 
It then seems unreasonable to insist that space-time remains classical at all scales.
It is more plausible that both space-time and  fields should have a 
unified quantum description in a fundamental theory.

This suggests the idea that space-time is not just a manifold 
which provides the stage for physics.
Rather, space-time ought to be {\em a dynamical physical system with intrinsic quantum structure},
which should be treated at the same footing as the fields %-- such as Yang-Mills gauge fields --
that live on it. This idea will be realized through a simple matrix model,  
where space-time arises as solution, and
gauge fields arise as fluctuations of the space-time structure.
In other words, space-time along with physical fields {\em emerge} from the basic matrix degrees of freedom.
This takes the idea of unification in physics one step further, and the new description 
is indeed simpler than the previous one(s), as in all successful 
unification steps in physics.

Of course this proposal is speculative and certainly not new, but it is
natural, given the special role of gravity as mediator between space-time and matter. 
The present specific model shows that it is in fact feasible, 
and it might be the key to solve some big puzzles in this context.
%The approach should not be viewed as holographic but rather as almost-local. 

We will also argue that an algebraic realization of such a space-time structure 
quite generally leads to some kind of higher-spin theory, if explicit breaking  
of rotational symmetry should be avoided. The reason is that a
 noncommutative structure of space-time amounts to some 
antisymmetric tensor field $\theta^{\mu\nu}$. Avoiding symmetry-breaking through 
some non-vanishing VEV $\langle\theta^{\mu\nu}\rangle \neq 0$
means that there should be a non-trivial variety of such objects at each point. This amounts to 
an internal structure of space-time, whose fluctuations lead to a higher-spin theory. 
A related idea was already discussed in \cite{Doplicher:1994tu}, without
making the connection to higher spin.
Such a structure is realized in a class of covariant
quantum spaces, whose prototype is fuzzy $S^4_N$ 
\cite{Grosse:1996mz,Castelino:1997rv,Ramgoolam:2001zx,Ho:2001as,Kimura:2002nq} in the Euclidean case. 
The model under consideration here \cite{Sperling:2019xar} provides
a minimal realization of  a covariant quantum space-time with Minkowski signature, and leads
to a consistent higher-spin gauge theory.

The present approach uses ideas of noncommutative (NC) geometry, but it  goes beyond it. 
NC geometry provides the insight that space-time does not need to be a classical manifold,
but can be described by some  algebra (of quantized ``functions'')  which need not be commutative. 
This is well-known by now and will be used throughout.
However, NC geometry  does not offer a model for a {\em dynamical} NC space-time.
This is exactly what matrix models provide in an extremely simple  way, leading
naturally to a gauge theory.
Upon closer inspection, such matrix models turn out to display stringy features.
This connection with string theory is not imposed by hand, but it is  useful to understand the UV 
properties of models \cite{Steinacker:2016nsc,Iso:2000ew}. In particular, loops involving such string-like modes generically
lead to a strong non-locality known as UV/IR mixing \cite{Minwalla:1999px}. This is avoided only in
a very specific and simple supersymmetric matrix model of the form 
\begin{align}
S[Y,\Psi] = {\rm Tr}\big( [Y^a,Y^b][Y_{a},Y_{b}]  \,\, + \overline\Psi \Gamma_a[Y^a,\Psi] \big) \ 
\label{MM-action}
\end{align}
involving 10 hermitian matrices $Y^a,\, a= 0,...,9$ and Majorana-Weyl fermionic matrices $\psi$.
This is  the well-known  IKKT or IIB model \cite{Ishibashi:1996xs}
which was proposed some 20 years ago as a constructive definition of string theory.
Having a preferred model is of course  welcome, and
analogous (and almost tantamount) to the statement that among all conceivable string theories,
there is 
basically only one anomaly-free model, which is critical (super)string theory in 10 dimensions.
%This will be explained in more detail section \ref{sec:quantization}.
Moreover, this model seems to provide just enough structure to possibly 
recover particle physics. 
 However that is very much aspirational, and  we refer to  \cite{Chatzistavrakidis:2011gs,Aschieri:2006uw,Sperling:2018hys,Aoki:2014cya} for work towards this goal.

Finally and perhaps most importantly, matrix models peg to be put on a computer. 
This is not easy in the  case of Minkowski signature, but large-scale efforts 
are underway to implement and simulate the full IKKT model \cite{Kim:2011cr,Nishimura:2019qal}, with tantalizing evidence towards an expanding 3+1-dimensional 
space-time structure. The long-term goal must then be to relate and test  analytical investigations as discussed here 
with such numerical results. This goal should be achievable, which provides a great  
motivation for more work in this direction. 

This article is basically a conceptual introduction and summary of the papers \cite{Sperling:2018xrm,Sperling:2019xar,Steinacker:2019dii,Steinacker:2019awe},
emphasizing the ideas behind it and minimizing the technicalities.
All the details can be found in the above papers, and  further supplementary literature 
is suggested in section \ref{sec:lit}.
There are also some new technical results, such as the regularization in 
\eq{MM-Mink} and \eq{path-integral}, and the treatment of diffeomorphisms 
in section \ref{sec:higher-spin-alg}.

%The outline of this article is as follows: ...

\section{The regularized IKKT or IIB model \& matrix geometry}
\label{sec:regularize}

The starting point of our considerations is the so-called IKKT or IIB matrix model \eq{MM-action},
which is an action for  hermitian matrices $Y^a \in \End(\cH), \ a=0,...,9$ acting on a Hilbert space $\cH$, and 
spinors $\Psi$ whose entries are (Grassmann-valued) matrices.
This model has a manifest $SO(9,1)$ symmetry, and is invariant under gauge transformations
\begin{align}
 Y^a \to U Y^a U^{-1}
 \label{gaugetrafo-matrix}
\end{align}
by arbitrary unitary matrices $U$ acting on $\cH$.
The model also enjoys maximal  supersymmetry  \cite{Ishibashi:1996xs}, which is important for the 
quantization but will not be used explicitly here.

\paragraph{From matrices to geometry.}

%We will largely ignore the fermions in this paper.
A priori, the model knows nothing about geometry, except for the 
$SO(9,1)$-invariant tensor $\eta_{ab}$ which enters the action. 
Obviously there is no way to assign any geometric significance to generic matrix configurations 
$\{Y^a\}$, and most of the configuration space is basically ``white noise''.
However, since the action is the square of commutators, it prefers 
matrix configurations which are almost-commutative. This would be obvious in the 
Euclidean signature case where $\eta_{ab}$ is replaced by $\d_{ab}$; consider this case for the moment.
Then the  dominant contributions are matrices which are almost-commutative, which
means that the $Y^a$ can ``almost'' be simultaneously diagonalized.
More specifically,  one can define quasi-coherent states $|y\rangle \in\cH$ 
which are optimally localized in a suitable sense\footnote{They are defined as  ground state(s) of $\sum_a(Y^a-y^a\one)^2$, and \eq{coherent-expect} is one way 
to detect the location of $\cM \subset \R^{9,1}$.} 
\cite{Ishiki:2015saa,Schneiderbauer:2016wub,Berenstein:2012ts}. These are the
approximate common eigenstates of the $Y^a$, localized at some point in target space
\begin{align}
 y^a = \langle y|Y^a|y\rangle \quad \in  \R^{9,1} \ .
 \label{coherent-expect}
\end{align}
These $y^a$ sweep out some variety\footnote{More precisely, one should consider the variety of coherent states $|y\rangle$, which may be 
degenerately embedded in target space via \eq{coherent-expect}. This is indeed what happens in section \ref{sec:M31}, 
leading to an internal bundle structure, 
cf. \cite{Karczmarek:2015gda,Sperling:2018xrm}.} in target space. 
This can even be implemented on a computer \cite{Schneiderbauer:2016wub}. 
One can thus associate classical functions to the matrices,
\begin{align}
 Y^a \sim y^a :\quad \cM \hookrightarrow \R^{9,1}
 \label{Y-embed}
\end{align}
which is interpreted geometrically as quantized embedding map of some ``brane'' $\cM$ in target space $\R^{9,1}$.
In this way, a fuzzy notion of geometry is extracted from  nearly-commuting matrix configurations. In the irreducible case 
the $Y^a$ will generate the full matrix algebra 
\begin{align}
  \End(\cH) \sim\cC(\cM) \ ,
\end{align}
which is interpreted as algebra of function on  $\cM$ in the spirit of noncommutative geometry  \cite{connes1994noncommutative}.
The brane may be a sharply defined submanifold, or it may be fuzzy in all directions.
It can also carry  internal extra dimensions which are not resolved by the $Y^a$.
In any case, one can then view the commutator
\begin{align}
 -i[Y^a,Y^b] =:\Theta^{ab} \sim \{y^a,y^b\} = \theta^{ab}
 \label{Poisson-bracket}
\end{align}
as an antisymmetric tensor field on $\cM$. More generally, $[.,.] \sim i \{.,.\}$
defines an anti-symmetric bracket on the space of ``fuzzy functions'' $ \End(\cH) \sim\cC(\cM)$,
which is a derivation and satisfies the Jacobi identity.
In other words, one can typically extract a Poisson bracket on $\cM$ in a low-energy limit, where the functions are
approximately commuting. An effective metric is encoded in the matrix Laplacian or d'Alembertian
\begin{align}
 \Box_Y \phi = [Y^a,[Y_a,\phi]], \qquad \phi \in \End(\cH) \ .
\end{align}
This is close in spirit to noncommutative geometry, however the extra ``embedding'' information \eq{Y-embed}
provides a useful and more 
direct access to the geometry, e.g. via coherent states.
A priori, it is not evident whether $y^a$ should be interpreted as space(time) coordinates,
or as some other coordinate functions on a higher-dimensional fuzzy phase space as in section \ref{sec:M31}.
This can be determined from the effective action describing the physics of  the fluctuations in the matrix model.

Now we return to the IKKT model.
Due to the Minkowski signature, the above argument seems to have a loop-hole, since there
can be configurations whose space-like commutators $[Y^i,Y^j]$ are large but canceled by 
the equally large space-time contributions $[Y^0,Y^i]$. 
%Such configurations clearly exist and may contribute significantly to the integral.
However, these are then solutions with high energy, as measured by
\begin{align}
 %T &:= \sum_{a,b} [Y^a,Y^b][Y^a,Y^b] \nn\\
 E &= T^ {00} = 2 [Y^0,Y^i][Y^0,Y_i] + \frac 12 [Y^a,Y^b][Y_a,Y_a] 
   = [Y^0,Y^i][Y^0,Y_i] + \frac 12[Y^i,Y^j][Y_i,Y_j]
\end{align}
for $i,j=1,...,9$.
Here $T^{ab}$ is the matrix energy momentum tensor  \cite{Steinacker:2008ri}, which 
satisfies $[Y_a,T^{ab}] = 0$.
As usual, the most fundamental and significant solutions of the model should be
those with lowest energy. 
Theses are then almost-commutative configurations according to the above discussion, 
which can be interpreted as quantized branes embedded in target space
with a semi-classical description as Poisson manifolds \eq{Poisson-bracket}.

\paragraph{Quantization and path integral, IR regularization.}

Perhaps the most important aspect of matrix models is that they provide a  
natural notion of quantization, by integrating over the space of all matrices.
For finite-dimensional hermitian matrices, the measure is the obvious one, 
which is invariant under the transformations \eq{gaugetrafo-matrix}.
Then  in the Euclidean case, the ``matrix path integral''
\begin{align}
 Z = \int dY e^{-S_E[Y]}  
 \label{path-integral-E}
\end{align}
is well-defined for traceless $Y^a$ \cite{Krauth:1998yu,Austing:2001pk}.
%even in the presence of flat directions $Y^a \to Y^a + c^a$.
In the case of Minkowski signature, the analogous integral is oscillating and not well-defined a priori.
However, it can be regularized. One possibility is to put an IR cutoff in both space-like
and time-like directions  as in \cite{Kim:2011cr}.
A similar but more elegant regularization is by adding a mass term 
$\Tr( m^2 Y^a Y^b \eta_{ab})$ to the model, and
giving the mass a suitable imaginary part.
We thus define 
\begin{align}
 S_\varepsilon[Y] &= \frac 1{g^2}\Tr \Big([Y^a,Y^b][Y_a,Y_b] - 2 m^2 \big(-e^{i\varepsilon}(Y^0)^2 + e^{-i\varepsilon}(Y^j)^2\big)\Big)\nn\\ 
  &\stackrel{\varepsilon \to 0}{\approx} \frac 1{g^2}\Tr \Big([Y^a,Y^b][Y_a,Y_b] - 2 m^2 Y^a Y_a \Big) 
 \label{MM-Mink}
\end{align}
where $j=1,...,D$ and $a,b=0,...,D$, which for $\varepsilon =0$ leads to the equations of motion (eom)
\begin{align}
 \Box_Y Y^a + m^2 Y^a = 0 \ . % \qquad \Box_Y = [Y^a,[Y_a,.]] \ .
 \label{eom}
\end{align}
Then the integral 
\begin{align}
 Z_\varepsilon = \int dY e^{i S_\varepsilon[Y]}  
 \label{path-integral}
\end{align}
(and similarly with fermions)
is absolutely convergent for $\varepsilon \in(0,\frac{\pi} 2)$,
at least for finite-dimensional matrices.
It turns out that 
this imposes at the same time Feynman's $i\varepsilon$ - prescription, so that 
\eq{path-integral} provides a solid definition for the quantized model. 
However, we will restrict 
ourselves to classical level here, and focus on the bosonic action \eq{MM-Mink} henceforth.
Note that  $m^2$ simply sets the scale of the theory, and
 there would be no scale in the model without mass term.

\section{Unification of space-time and gauge fields}

The cubic matrix equation \eq{eom} has many different solutions with very different significance.
Finding the ``dominant'' one(s) is a non-perturbative problem which we will not address here, and
we will simply choose some solution which leads to interesting physics.
Whatever background we choose, the fluctuations in the matrix model automatically 
defines a gauge theory.
Indeed if $X^a$ is some background solution, then the fluctuations 
\begin{align}
 Y^a = X^a + \cA^a, \qquad \cA^a \in \End(\cH)
 \label{covar-coords}
\end{align}
around this background are parametrized in terms of tangential (and possibly transversal) 
 modes $\cA^a$, interpreted as vector fields on $\cM$. 
They transform under gauge transformations as
\begin{align}
 \cA_a \to U^{-1} \cA_a U + U^{-1}[X_a,U] \ .
\end{align}
Since $[X_a,.]$ is a derivation,
this clearly corresponds to the inhomogeneous transformation law for 
gauge fields in a Yang-Mills-type gauge theory, whose precise form depends 
on the background. This will be discussed briefly for the Moyal-Weyl solution $\R^{3,1}_\theta$ below, and
worked out in more detail for the fuzzy space-time $\cM^{3,1}_n$ in section \ref{sec:modes}.

Note that \eq{covar-coords} suggests an interpretation of the gauge fields $\cA^a$  as Goldstone bosons of the 
spontaneously broken translational symmetry\footnote{In the presence of some potential such as $Y_a Y^a$, the $SO(9,1)$ rotations 
play the role of this symmetry.} $Y^a \to Y^a + c^a \one$ of the matrix model.
These modes are accordingly massless, as they should be.
%This is yet another new insight gained in the framework of matrix models. 

\section{Examples of matrix geometries}
\label{sec:examples}

In this section we discuss two basic examples of embedded noncommutative spaces
described by finite or infinite matrix algebras. The salient feature is that the geometry is defined by a specific set of 
matrices $X^a$, interpreted as quantized embedding maps of a sub-manifold in $\R^D$.
We will learn to freely switch between the noncommutative matrix setting and the 
semi-classical picture of Poisson manifolds.

\subsection{Prototype: the fuzzy sphere $S^2_N$}

As a first example, we recall the
fuzzy sphere  $S^2_N$  \cite{Madore:1991bw,hoppe1982QuaTheMasRelSurTwoBouStaPro}. This is a 
quantization or matrix approximation of the usual sphere $S^2$,
with a cutoff in  angular momentum.
The starting point is the observation that the algebra of functions $\cC(S^2)$ on $S^2$
is generated by the Cartesian coordinate functions $x^a$ of $\R^3$ modulo the
relation $ \sum_{ {a}=1}^{3} {x}^{a}{x}^{a} = 1$. 
Similarly, $S^2_{N}$ is a
non-commutative space defined in terms of three $N \times N$ hermitian matrices $X^a,\ a=1,2,3$ 
subject to the relations
\begin{equation}
[ X^{{a}}, X^{{b}} ] = \frac{i}{\sqrt{C_N}}\varepsilon^{abc}\, X^{{c}}~ , 
\qquad \sum_{{a}=1}^{3} X^{{a}} X^{{a}} =  \one 
\end{equation}
where $C_N= \frac 14(N^2-1)$ is the value of the quadratic Casimir of 
$\msu(2)$ on $\cH = \C^N$.
They are realized by the generators of the $N$-dimensional irrep $(N)$ of $\msu(2)$. 
The matrices $X^a$ should be interpreted as quantized embedding functions
in the Euclidean target space $\R^3$,
\begin{align}
 X^a \sim x^a:\quad S^2 \hookrightarrow \R^3.
\label{embedding-S2}
\end{align}
They generate the matrix algebra $\End(\cH)$,
which is viewed as quantized algebra of functions on the symplectic space $(S^2,\omega_N)$.
Here $\omega_N$ is the  $SU(2)$-invariant symplectic form on $S^2$
with $\int \omega_N = 2\pi N$. 
This is best understood by decomposing $\End(\cH)$  into irreps of the adjoint action of $SU(2)$, 
\begin{align}
\End(\cH) \cong (N) \otimes (\bar N) 
&= (1) \oplus (3) \oplus ... \oplus (2N-1) \nn\\
&= \{\hat Y^{0}_0\} \,\oplus \, ... \, \oplus\, \{\hat Y^{N-1}_m\}.
\label{fuzzyharmonics}
\end{align}
This provides the definition of the 
fuzzy spherical harmonics $\hat Y^{l}_m$, which are symmetric traceless polynomials in $X^a$
of degree $l$.
It also provides the  $SO(3)$-invariant {\em quantization map}\footnote{The normalization 
can be fixed by requiring that $\cQ$ respects the norm defined via \eq{integral-S2}.}
\begin{align}
\begin{array}{rcl}
\cQ: \quad \cC(S^2) &\to& \,\, \End(\cH)\, \\
 Y^l_m &\mapsto& \left\{\begin{array}{c}
                         \hat Y^l_m, \quad l<N \\ 0, \quad l \geq N \ .
                        \end{array}\right.
\end{array}
\label{quant-map-S^2} 
\end{align}
One can easily verify $\cQ(x^a) = X^a$ and $\cQ(i\{x^a,x^b\}) = [X^a,X^b]$ where $\{,\}$ denotes the Poisson brackets 
corresponding to the symplectic form $\omega_N = \frac N2 \varepsilon_{abc} x^a dx^b dx^c$ on $S^2$, and more generally 
\begin{align}
\frac 1N\big(\cQ(i\{f,g\}) -[\cQ(f),\cQ(g)]\big) \stackrel{N\to\infty}{\to} 0 \ .
\end{align}
%Hence $S^2_N$  is the quantization of $(S^2,\omega_N)$.
Furthermore, the following integral relation holds
\begin{align}
 \Tr (\cQ(f)) = \int\limits_{S^2} \frac{\omega_N}{2\pi} f \ .
 \label{integral-S2}
\end{align}
This means that $S^2_N$ is  the  quantization of $(S^2,\omega_N)$. The correspondence is summarized 
in table \ref{tab:correspondence}, and
an analogous dictionary applies to all fuzzy spaces considered here.
Optimally localized coherent states $|x\rangle$ are given by $SU(2)$ rotations
of the lowest weight state of $\cH$.
\begin{table}[h]
\begin{center}
 \begin{tabular}{c|c}
  noncommutative/fuzzy space $\cM_n$ & semi-classical space $\cM$  \\ \hline\hline
$\End(\cH) \ \ni \hat\phi(X)  $ & $ \cC(\cM)  \ \ni \phi(x) $  \\  \hline
$X^a$ & $x^a$  \\   \hline
$[\hat\phi,\hat\psi]$ &  $i\{\phi,\psi\}$  \\ \hline
$\Tr \hat\phi(X)$ &  $ \int d\Omega \phi(x)$  \\ \hline
 \end{tabular}
\caption{Schematic correspondence between matrices (operators) in $\End(\cH)$ 
and functions on $\cM$.
$d\Omega$ indicates the symplectic volume form. The metric structure is encoded 
in the Laplacian $\Box$.}
\label{tab:correspondence}
\end{center}
\end{table}
The (round) metric is encoded in  the  $SU(2)$-invariant Laplacian
\begin{align}
 {\Box} = [X^a,[X^b,.]]\d_{ab} \ ,
\label{matrix-laplacian-S2}
\end{align}
which is nothing but the 
quadratic $SU(2)$ Casimir on $S^2$. It is  easy to see that  
its spectrum coincides with the spectrum of the classical 
Laplace operator on $S^2$ up to the cutoff, and the eigenvectors are given by the fuzzy spherical harmonics $\hat Y^l_m$.
Finally, $S^2_N$ is a solution of  \eq{eom} with $\Box X^a = 2 X^a$.

\subsection{The Moyal-Weyl solution $\R^{3,1}_\theta$ and $\cN=4$ SYM}

If we allow infinite-dimensional matrices, the model can also describe non-compact brane solutions, 
including the much-studied Moyal-Weyl quantum plane $\R^{3,1}_\theta$. 
This is defined in terms of 
operators $X^\mu\in\End(\cH)$ which satisfy 
\begin{align}
 [X^\mu,X^\nu]  = i \theta^{\mu\nu}\one, \qquad  \mu=0,...,3
\end{align}
where $\theta^{\mu\nu}$ is a constant anti-symmetric tensor. 
The analog of the quantization map \eq{quant-map-S^2} is given by Weyl quantization,
or equivalently as 
\begin{align}
 \cQ: \quad \cC(\R^{3,1}) &\to End(\cH)  \nn\\
 \phi(x)  &\mapsto \hat \phi := \int\limits_{\R^{3,1}} \phi(x) 
\left|x\right\rangle \left\langle x\right| \ 
 \label{quantization-map-1}
\end{align}
with inverse map $\hat\phi \to \phi(x) =\langle x |\hat\phi|x\rangle$,
where $|x\rangle$ are the  coherent states.
This is a solution of the massless model \eq{eom} with $\Box X^\mu = 0$.
Parametrizing the fluctuations
around this background as\footnote{These fluctuating noncommutative coordinates are aptly called 
``covariant coordinates'' in NC field theory  \cite{Madore:2000en}.} 
$Y^\mu = X^\mu+\cA^\mu(X)$ and $Y^i = \phi^i(X), \ i=4,...,9$ and
including the fermions of the IKKT model, one recovers  
$\cN=4$ NC super-Yang-Mills theory (SYM) on $\R^{3,1}_\theta$ \cite{Aoki:1999vr}.
This is best seen by recalling that usual $\cN=4$ SYM is obtained by dimensional reduction of 
$\cN=1$ SYM on $\R^{9,1}$ to $\R^{3,1}$, while the IKKT model is nothing but $\cN=1$ SYM on $\R^{9,1}$ dimensionally reduced to 
a point; hence the basic structure is the same, 
and the covariant derivatives on $\R^{3,1}_\theta$  arise 
 via $[X^\mu+\cA^\mu,.] = i D_\mu$.

An important feature  is that  the UV divergences are 
canceled due to the extended supersymmetry, which entails that also the notorious IR divergences are absent, 
which usually plague NC field theory. 
Note that this is a property of the model rather than the background; the cancellation of divergences
happens on any 4-dimensional background, as discussed in \cite{Steinacker:2016nsc} and in section \ref{sec:quantization}.
In this respect the model is pretty much unique \cite{Matusis:2000jf,Jack:2001cr}, 
and that is the reason for focusing on it.

However, there is an issue: %while translation invariance is respected,
the explicit $\theta^{\mu\nu}$ breaks Lorentz invariance. Even though it
is not manifest in the action, loop corrections may lead to significant 
Lorentz-violating effects. For example, the induced 
gravity action  will contain unwanted terms such as 
$\int R_{\mu\nu\a\b}\theta^{\mu\nu}\theta^{\a\b}$, 
where $R_{\mu\nu\a\b}$ is the Riemann tensor of the 
effective metric \cite{Klammer:2008df}, which is  dynamical \cite{Rivelles:2002ez,Steinacker:2010rh}. 
This issue is resolved on {\em covariant quantum spaces}, where 
the fixed $\theta^{\mu\nu}$ is replaced by a 
bundle-like variety of different $\{\theta^{\mu\nu}\}$ over space-time,
which is effectively averaged.
Thus space-time acquires a non-trivial internal structure,
which  restores Lorentz invariance at least partially. 
It also means that the  fluctuation modes will  involve
harmonics on this internal space, leading to a {\em higher-spin theory}.

\section{Fuzzy $H_n^{4}$ and higher spin}
\label{sec:fuzzyH4}

As a basis for the quantum space-time discussed in section \ref{sec:M31}, we now discuss a 
more sophisticated solution of the model
known as fuzzy 4-hyperboloid $H^4_n$. This is 
a prototype of a covariant quantum space with an interesting internal structure, which
nicely illustrates the ideas discussed in the introduction leading 
to a higher-spin gauge theory. 
The starting point is the Lie algebra $\mso(4,2)$
generated by $M^{ab}$,  
\begin{align}
  [M_{ab},M_{cd}] &= i \left(\eta_{ac}M_{bd} - \eta_{ad}M_{bc} - 
\eta_{bc}M_{ad} + \eta_{bd}M_{ac}\right) \ 
 \label{M-M-relations-noncompact}
\end{align}
where  $\eta_{ab} = {\rm diag}(-1,1,1,1,1,-1)$ and  $a,b=0,..,5$,
and a specific class of
unitary representations  $\cH_n$ known as doubletons or 
minireps \cite{Mack:1975je,Fernando:2009fq}, labeled by  $n\in\N$. 
These are short discrete series unitary irreps of 
$\mso(4,2)$, which are   lowest weight representations that are multiplicity-free
and remain irreducible if restricted to $SO(4,1)\subset SO(4,2)$. 
In particular, unitarity means that all $M^{ab}$ are hermitian operators.
%The  case $n=0$ is excluded.
We  define 
the fuzzy hyperboloid $H^4_n$ \cite{Hasebe:2012mz,Sperling:2018xrm} in terms of hermitian operators   
\begin{align}
 X^a := r M^{a5}, \qquad a=0,...,4 \ .
\end{align}
%%H V3 added
which are viewed as quantized embedding functions of a brane in target space,
\begin{align}
 X^a \sim x^a: \quad H^4 \ \hookrightarrow \ \R^{4,1} \ .
 \label{H4-embedding}
\end{align}
Here $r$ has dimension of length. 
Since $\cH_n$ is irreducible under $SO(4,1)$, they satisfy
the relations of a 4-dimensional Euclidean hyperboloid
\begin{align} 
 \eta_{ab} X^a X^b &=  - R^2 \one \  ,\qquad
 R^2 = \frac{r^2}{4}(n^2-4) \ .
  \label{X-constraint}
\end{align}
Even though this suggests that the noncommutative space described by the $X^a$ is just a hyperboloid,
this is not quite correct. To understand this construction, we first note that 
$X^a$ generate the full algebra $\End(\cH_n)$, because their commutators 
\begin{align}
 [X^a,X^b] = -i r^2 M^{ab} =: i \Theta^{ab}
 \label{X-X-CR}
\end{align}
are nothing but the generator of $\mso(4,1)$ on $\cH_n$. 
This is analogous to the case of Snyder space \cite{Snyder:1946qz}.
Since the fluctuations $X^a + \cA^a$ in \eq{covar-coords} are the most general elements in $\End(\cH_n)$, 
we must find the proper interpretation of $\End(\cH_n)$ as a quantized algebra of functions on some space.
To understand this, we note that $\End(\cH_n)$ transforms under $SO(4,2)$ via
\begin{align}
 M^{ab} \triangleright\phi = [M^{ab} ,\phi]  \ , \qquad \phi \in \End(\cH_n) \ , \qquad  a,b=0,...,5
\end{align}
while the $X^a$ transform covariantly as vector operators of $SO(4,1)\subset SO(4,2)$,
\begin{align}
 M^{ab} \triangleright X^c = [M^{ab} ,X^c] = i(\eta^{ac} X^b - \eta^{bc} X^a)\ , \qquad  a,b=0,...,4
 \label{covariance}
\end{align}
This strongly suggests that $\End(\cH_n)$ is the quantized algebra of functions on some coadjoint orbit of 
$SO(4,2)$ or $SU(2,2)$, which turns out to be quantized $\C P^{1,2}$. This is an $S^2$ -bundle over $H^4$, whose fiber is given by 
the space of selfdual 2-forms $\theta^{ab}$ on $H^4$.
\begin{figure}[h]
\centering
 \includegraphics[width=0.2\textwidth]{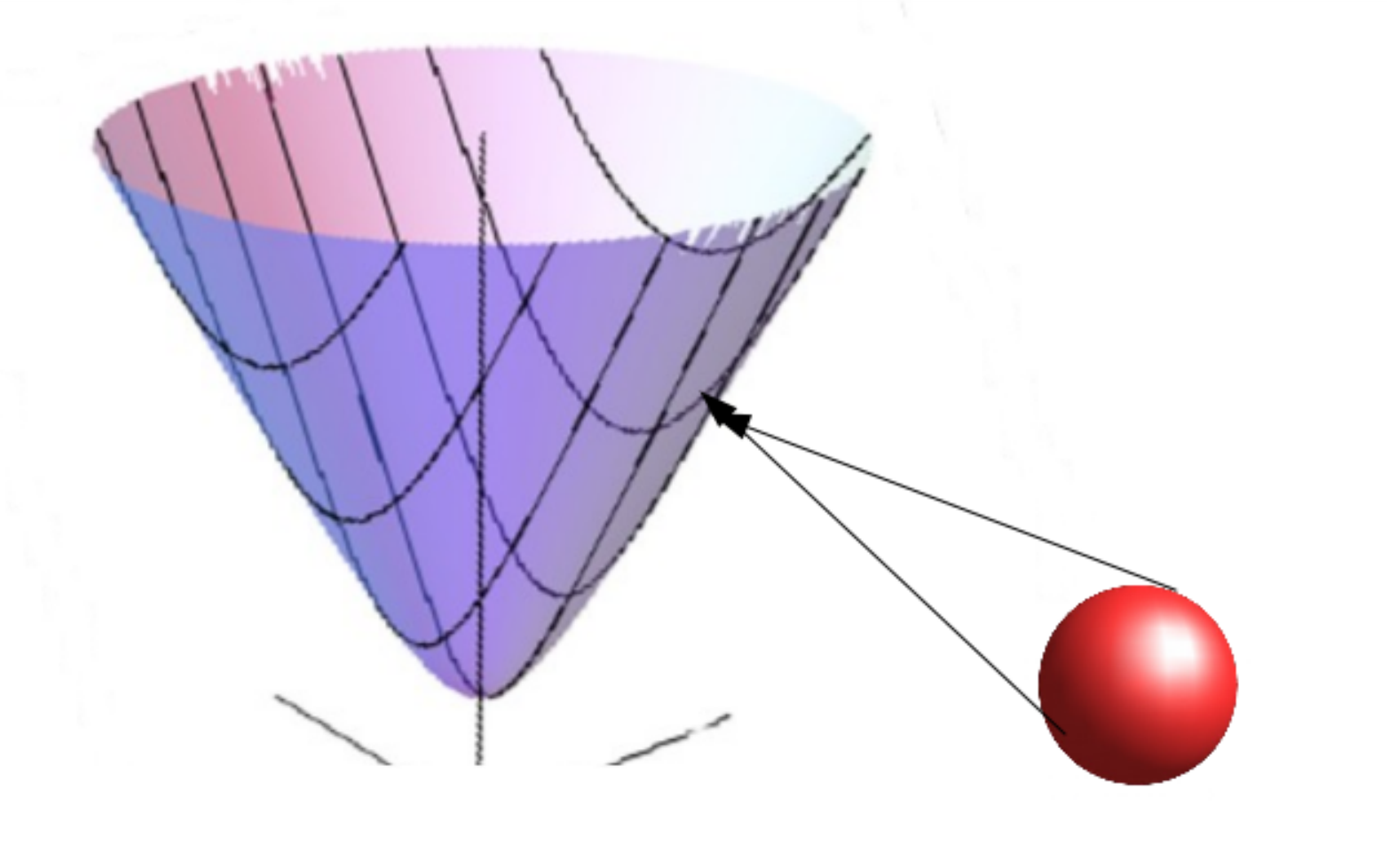}
\end{figure}
Covariance means that the bundle is an $SO(4,1)$-equivariant bundle, which means that 
the action on the bundle is compatible with the action on the base manifold. 
Thus the local stabilizer $SO(4)$ of any point $p\in H^4$ acts non-trivially on the 
$S^2$ fiber over $p$. This  implies that the ``would-be Kaluza Klein''  modes on $S^2$ will transmute into 
higher spin modes on $H^4$, leading to a higher-spin gauge theory on $H^4$ in the matrix model.
Because this is a crucial aspect, we briefly sketch how this bundle arises:

\paragraph{Bundle construction.}

This bundle is obtained explicitly 
via an oscillator construction, which at the same time provides the doubleton representations $\cH_n$.
Consider 4 bosonic oscillators $\psi^\a,\ \a=1,...,4$ which satisfy 
\begin{align}
 [{\psi}_\a, \obar{\psi}^\b] = \d_\a^\b
\end{align}
where $\obar{\psi} = \psi^\dagger \g_0$ with  $\gamma_0 = diag(1,1,-1,-1)$.
They transform in the fundamental representations $\C^4$ of $SU(2,2)$. Then consider 
the Jordan-Schwinger realization of $\msu(2,2) \cong \mso(4,2)$ via 
\begin{align}
 M^{ab} &=\obar{\psi}\Sigma^{ab}\psi, \qquad   a,b=0,...,5  \\[1ex]
 X^a &= \frac r2 \obar{\psi}\gamma^{a}\psi, \qquad   a=0,...,4 
 \label{M-X-bundle-def}
\end{align}
where $\Sigma^{ab}$ are the generators of $\mso(4,2)$ on $\C^4$,
noting that $\Sigma^{a5} = \frac 12 \g^a$.
The $M_{ab}$ are  hermitian, and thus implement unitary representations of $SU(2,2)$ on the 
Fock space of the bosonic oscillators. 
More precisely, one defines 
\begin{align}
\psi_\a = (a^\dagger_{1},  a^\dagger_{2}, b_1, b_2)^T \ ,
\label{spinor-oscillator}
\end{align}
and considers the  Fock vacuum 
$a_i \left|0\right\rangle = 0 = b_i \left|0 \right\rangle$. Then 
the doubleton minireps $\cH_n$ arise on the lowest weight vectors
\begin{align}
\begin{aligned}
  \left|\Omega \right\rangle &:= 
  \left|E,\frac{n}{2},0 \right\rangle &:= 
a_{i_1}^\dagger \ldots a_{i_n}^\dagger \left|0 \right\rangle, 
 \qquad E = 1+\frac{n}{2},\ j_L = \frac n2, \  j_R = 0 
 \end{aligned}
 \label{minireps-Fock}
\end{align}
which satisfy $a_i b_j |\Omega\rangle = 0$.
Similarly, exchanging $a$ with $b$ exchanges $j_L \leftrightarrow j_R$.
Since the  ``norm'' 
\begin{align}
 \obar{\psi}\psi = -N_a + N_b - 2 
\end{align}
is  invariant under $SU(2,2)$, 
the $\psi_\a$ clearly form a noncompact version of $\C P^3$ in the (semi)classical limit, which is called $\C P^{1,2}$.
This space is in fact a coadjoint orbit of $SU(2,2)$, and thus carries a canonical symplectic form
corresponding to the Poisson structure 
\begin{align}
  \{m^{ab},m^{cd}\} &=  \eta^{ac}m^{bd} - \eta^{ad}m^{bc} - \eta^{bc}m^{ad} + \eta^{bd}m^{ac} \nn\\
  \{x^a,x^b\} &= \theta^{ab}  = -r^2 m^{ab}
 \label{M-M-Poisson}
\end{align}
considering $m^{ab}\sim M^{ab}$ as embedding functions 
\begin{align}
m^{ab}: \ \ \C P^{1,2} \ \hookrightarrow  \ \msu(2,2) \cong \R^{15} \ .
\end{align}
The semi-classical limit is indicated by $\sim$.
In particular, we note that the definition of $X^a$ in \eq{M-X-bundle-def} amounts to a quantized Hopf map from the bundle space to $H^4$,
which is compatible with $SO(4,1)$. The radius of $H^4$ is related to the quantum number $n\in\N$ \eq{minireps-Fock} via 
\begin{align}
 X_a X^a =  - \frac {r^2}4 (n^2-4) =: - R^2 \ .
 \label{R2-fuzzy-explicit}
\end{align}
It can also be checked that these $\theta^{ab}$ are self-dual 2-forms on $H^4$ in the semi-classical limit, which 
sweep out the $S^2$ fiber over $H^4$.
For more details we refer to \cite{Govil:2013uta,Sperling:2018xrm}.
One can also see  that the local $SO(4) \cong SU(2)_L \times SU(2)_R$ stabilizer e.g. over $x^a = (r,0,0,0,0) \in H^4$ acts 
via $SU(2)_L$ on the 
2-component spinor $(a_1,a_2)\in\C^2$, sweeping out $\C P^1 \cong S^2$. 
In the noncommutative case, this realizes precisely the oscillator realization of fuzzy 2-sphere $S^2_n$, which
will lead to a truncation of the higher spin modes in \eq{EndH-Cs-decomposition}.

In particular, we have gained a crucial understanding of $\End(\cH_n)$ as quantized algebra of functions on $\C P^{1,2}$,
which is an $S^2$ bundle over $H^4$. This can be made more explicit in terms of coherent states $|p\rangle \in\cH_n$,
which are optimally localized states at $p\in\C P^{1,2}$ given by $SO(4,2)$-rotations of $|\Omega\rangle$, 
cf. \cite{Steinacker:2016nsc}. One can then write down an 
$SO(4,2)$ -covariant quantization map analogous to \eq{quantization-map-1},
\begin{align}
 \cQ: \quad \cC(\C P^{1,2}) &\to End(\cH_n)  \nn\\
 \phi(p)  &\mapsto \hat \phi := \int\limits_{\C P^{1,2}} \phi(p) 
\left|p\right\rangle \left\langle p\right| \ 
 \label{quantization-map}
\end{align}
with inverse map  $\hat\phi \mapsto \langle p| \hat\phi|p\rangle= \phi(p)$ below some cutoff depending on $n$. In other words,
\begin{align}
\boxed{ \ 
\End(\cH_n) \ \cong \ \cC(\C P^{1,2})
\ } \qquad \mbox{up to some cutoff}
\end{align}
as vector spaces and $SO(4,2)$ modules.
More precisely, the Hilbert-Schmid operators in $\End(\cH_n)$ correspond to 
square-integrable functions on $\C P^{1,2}$, and both sides decompose into principal series irreps of $SO(4,2)$.
However \eq{quantization-map} applies more generally e.g. also to polynomial functions, which of course do not form unitary representations.

Finally, it is interesting to note that  $\C P^{1,2}$ and its quantization 
can be considered as (quantized) twistor space. However its present use is quite distinct from the usual 
applications of twistors.

\subsection{Algebraic description}

In the semi-classical limit, the above generators satisfy the following constraints 
\begin{align}
  x_a x^a &= -R^2, \label{eq:xx=R} \; \\
 \theta^{ab} x_b &= 0  \; ,\\
 \epsilon_{abcde} \theta^{ab} x^{c} &= n r\theta_{de} \sim 2 R \, \theta_{de}     
 \; ,  \label{SD-H-class}\\
  \eta_{aa'}\theta^{ab} \theta^{a'b'} &= \frac{L^4_{NC}}{4}   P^{bb'}\; ,
 \label{theta-constraint}
 \end{align}
  where  the scale of non-commutativity is
\begin{align}
 L^4_{NC} &:= \theta^{ab}\theta_{ab} = 4 r^2 R^2 \ .
 \label{LNC-def}
\end{align}
Here
 \begin{align}
 P^{ab} = \eta^{ab} + \frac{1}{R^2}x^a x^b  
 \qquad \text{with} \quad 
 P^{ab} x_b = 0  
\end{align}
is the Euclidean projector on $H^4$
(recall that $H^4$ is a Euclidean space).
Hence functions on fuzzy $H^4_n$ can be identified
for large $n$ with functions on  $\C P^{1,2}$, and written
in the form
\begin{align}
 \cC(\C P^{1,2})
 = \bigoplus\limits_{s=0}^\infty \cC^s  \quad \ni \ 
 \phi^s_{a_1\ldots a_s;b_1 \ldots b_s}(x) \ \theta^{a_1b_1} \ldots \theta^{a_sb_s} \ .
  % \ \equiv \phi^s_{\und{\b}}(x)\, \Xi^{\und{\b}}
 \label{C-module-semiclass}
\end{align}
This can be viewed as function on $H^4$ taking values in Young diagrams
${\tiny \Yvcentermath1 \young(aaa,bbb) }$.
\eq{C-module-semiclass} gives  a 
 decomposition of $\cC = \cC(\C P^{1,2})$ into modules $\cC^s$
over the algebra of functions $\cC^0$ on $H^4$, which correspond to bundles 
 over $H^4$ whose structure is determined by the above constraints\footnote{Another 
 description is given by the
one-to-one map
\begin{align}
    \Gamma^{(s)} H^4 \  &\rightarrow  \    
\cC^s  \nn\\
  \phi^{(s)}_{a_1 \ldots  a_s}(x)   &\mapsto  \ 
 \phi^{(s)} =  \{x^{a_1},\ldots\{x^{a_s},\phi^{(s)}_{a_1\ldots a_s}\}\ldots\} 
\ .
 \label{psi-iso}
\end{align}
Here $\Gamma^{(s)}H^4$ denotes the space of totally symmetric,
traceless, divergence-free rank $s$ tensor fields on $H^4$, which are identified  
with (symmetric tangential divergence-free traceless)
tensor fields $\phi^{(s)}_{a_1 \ldots  a_s}$ with  $SO(4,1)$ indices.}.
In the NC case, this decomposition is defined as 
\begin{align}
    \End(\cH_n) =\cC = \cC^0 \oplus \cC^1 \oplus \ldots \oplus  \cC^n\qquad \text{with} 
\quad 
  \cS^2|_{\cC^s} = 2s(s+1)  \, 
  \label{EndH-Cs-decomposition}
\end{align} 
in terms of the Casimir 
\begin{align}
\cS^2 &:= \frac 12\sum_{a,b\neq 5} [M_{ab},[M^{ab},.]] 
  + r^{-2} [X_a,[X^a,.]]  \
  = 2 C^2[\mso(4,1)] - C^2[\mso(4,2)]
\label{Spin-casimir}
 \end{align}
which can be interpreted as a spin observable on $H^4_n$. It satisfies
 \begin{align}
 [\cS^2,\Box_H] = 0    \ , \qquad \Box_H = [X^a,[X_a,.]] \ ,
\end{align}
and it is easy to check that $X^a$ is a solution of the eom \eq{eom},
\begin{align}
 \Box_H X^a = -4 r^2 X^a \ .
 \label{Box-H4n}
\end{align} 
Note that \eq{EndH-Cs-decomposition} has a cutoff at spin $s=n$,
which results from the fact that the fiber on fuzzy $\C P^{1,2}$ is really a fuzzy sphere $S^2_n$,
which supports only harmonics up to spin $n$. This is discussed briefly below \eq{R2-fuzzy-explicit} 
and shown in detail in \cite{Sperling:2018xrm}.
This cutoff disappears in the semi-classical limit, where
the  $\cC^s$  are modules over $\cC^0 \cong \cC(H^4)$, interpreted  as 
sections of higher spin bundles over $H^4$.

Even though this $H^4$ solution is ultimately unphysical,
the common features with Vasiliev's higher spin theory \cite{Vasiliev:1990en,Didenko:2014dwa}
 are most visible in this background, which is nothing but Euclidean $AdS^4$. 
Moreover it is the starting point for the quantum space-time $\cM^{3,1}_n$, which is discussed next.

\section{Cosmological space-time $\cM^{3,1}_n$}
\label{sec:M31}

Now we make a big step towards real physics, and discuss 
a solution of the model that describes a 
quantized cosmological FLRW space-time $\cM^{3,1}$. 
The fluctuation modes on this background lead to a consistent higher-spin
gauge theory, which is very interesting from the physics point of view.
We will review the basic definition of this solution \cite{Sperling:2019xar} and recent results,  including the 
no-ghost theorem \cite{Steinacker:2019awe} and the linearized Schwarzschild solution \cite{Steinacker:2019dii}. 

%%H new V3
Before giving the explicit mathematical realization, we briefly explain the idea.
As illustrated in the previous examples, noncommutative or quantum geometries are described by two 
structures: one is a (noncommutative) {\em algebra}, interpreted as quantized algebra of 
functions $\cC^\infty(\cM)$ of a classical manifold. This algebra encodes the abstract manifold,
and it is always $\End(\cH)$ in the present framework, for some separable Hilbert space $\cH$.
In addition we need to define a {\em metric} structure, corresponding to a Riemannian or Lorentzian manifold.
This is defined here through a matrix Laplacian or d'Alembertian\footnote{Alternatively one could use a Dirac operator, 
as in Connes axiomatic approach \cite{connes1994noncommutative}.} $\Box$ which acts on $\End(\cH)$, as in
\eq{matrix-laplacian-S2} and \eq{Box-H4n}. Thus the same algebra $\End(\cH)$ can describe very different geometries, 
even with different signatures, as the same abstract manifold can have different metrics.

\begin{figure}[h]
%\begin{center}
\hspace{3cm} \includegraphics[width=0.45\textwidth]{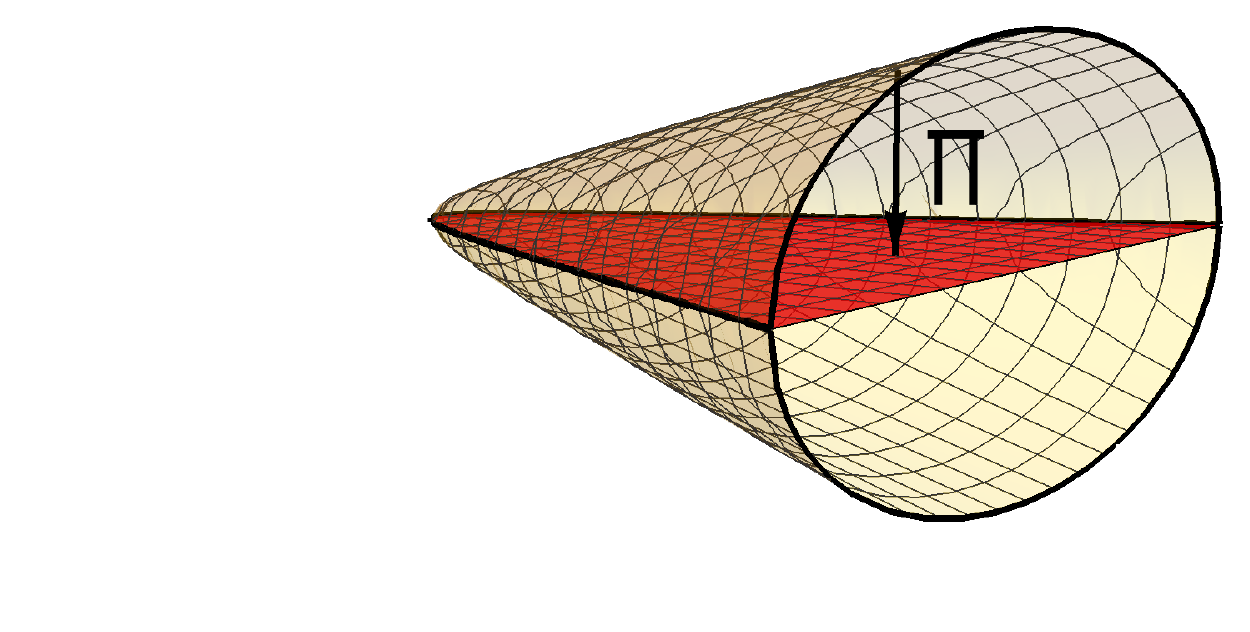}
 \caption{Projection $\Pi$ from $H^4$ to $\cM^{3,1}$  with 
Minkowski signature.}
%\end{center}
 \label{fig:projection}
\end{figure}

In this spirit, the  cosmological space-time $\cM^{3,1}_n$  under consideration 
coincides with the fuzzy hyperboloid $H^4_n$ as a (quantized) manifold, but it inherits a Lorentzian effective metric through 
a different matrix d'Alembertin $\Box$  \eq{dAlembert-fuzzy-def}, which 
governs the fluctuations around the background solution \eq{background-solution-M} in the matrix model.
An intuitive picture of $\cM^{3,1}_n$ is obtained in terms of quantized embedding functions 
from the manifold into target space, as in the case of fuzzy $S^2_N$ \eq{embedding-S2} and fuzzy $H^4_n$ \eq{H4-embedding}.
More specifically, we consider  {\em four} generators $X^\mu$ of $H^4_n$
as quantized embedding functions into target space
\begin{align}
 X^\mu \sim x^\mu: \quad \cM^{3,1} \ \hookrightarrow \ \R^{3,1}, \qquad  \ \mu = 0,...,3 \ ,
 \label{M31-embedding}
\end{align}
where Greek indices $\mu,\nu$ will run from $0$ to $3$ from now on.
This can be interpreted as a brane $\cM^{3,1}$ embedded in  $\R^{3,1}$.
Dropping $X^4$ means that the same abstract manifold is now embedded in $\R^{3,1}$ (rather than $\R^{4,1}$ 
as for $H^4_n$), so that $\cM^{3,1}_n$ can be interpreted as squashed hyperboloid
projected  to $\R^{3,1}$,
as sketched in figure \ref{fig:projection}. 
Accordingly, \eq{R2-fuzzy-explicit} is rewritten as
\begin{align}
 \eta_{\mu\nu} X^\mu X^\nu &=  - R^2 \one - X_4^2 \ \leq  - R^2 \one \ .
\end{align}
It is easy to see that the $X^\mu$ alone generate the full algebra $End(\cH_n)$
of (quantized) functions on $\C P^{1,2}$, which is now viewed as
quantized algebra of functions on an $S^2$-bundle over $\cM^{3,1}$. The 
$X^\mu\sim x^\mu$ define an embedding of this bundle in $\R^{3,1}$ which is degenerate along the fibers.
Although we will largely focus on the 3+1-dimensional base manifold, $\cM^{3,1}$ is understood to carry this bundle structure.

The above picture strongly suggests that  $\cM^{3,1}$ will carry an effective metric with Lorentzian signature.
That metric is encoded in the matrix d'Alembertian
\begin{align}
 \Box  \phi \equiv  \Box_T  \phi  &=  [T_\mu,[T^\mu,\phi]] =  (C^2[\mso(4,1)] - C^2[\mso(3,1)])\phi \ ,
 \label{dAlembert-fuzzy-def}
\end{align}
where 
 \begin{align}
 T^\mu = \frac{1}{R} M^{\mu 4} \ .
\end{align}
This defines an  $SO(3,1)$-invariant d'Alembertian for $\cM^{3,1}$
 with Lorentzian structure, which will govern all modes in the matrix model as discussed in section \ref{sec:fluctuations}.
Our task is then to describe the spectrum of fluctuations and their interactions. 
Note that we now have two vector generators, which satisfy the $SO(3,1)$ -covariant commutation relations
\begin{align}
 [X^\mu,X^\nu] &=: i\Theta^{\mu\nu} = -i r^2 M^{\mu\nu} \nn \\
 [T^\mu,T^\nu] &= -\frac{i}{r^2R^2}\Theta^{\mu\nu}, \nn\\
 [T^\mu,X^\nu] &= \frac{i}{R} \eta^{\mu\nu}\, X_4 \ .
  \label{XT-CR}
\end{align}
Both $X^\mu$ and $T^\mu$ provide solutions of the eom \eq{eom} for different signs of the mass term, 
\begin{align}
  \Box_X X^\mu &= -3 r^2\, X^\mu,  \qquad 
  \Box_T T^\mu = 3 R^{-2}\,  T^\mu \ .
  \label{Box-X-T}
\end{align}
These generators satisfy further constraints due to the special representation $\cH_n$,
which are crucial for the consistency of the resulting gauge theory.
To simplify these relations we will focus on the semi-classical (Poisson) 
limit $n \to \infty$ from now on, working with commutative functions of $x^\mu\sim X^\mu$ and $t^\mu \sim T^\mu$, 
but keeping the Poisson structure  $[.,.] \sim i \{.,.\}$ encoded in $\theta^{\mu\nu}$.

\paragraph{Semi-classical structure.}

In the semi-classical limit, the generators $x^\mu$ and $t^\mu$ satisfy 
the following constraints   \cite{Sperling:2018xrm}
\begin{subequations}
\label{geometry-H-M}
\begin{align}
 x_\mu x^\mu &= -R^2 - x_4^2 = -R^2 \cosh^2(\eta) \, , 
 \qquad R \sim \frac{r}{2}n   \label{radial-constraint}\\
 t_{\mu} t^{\mu}  &=  r^{-2}\, \cosh^2(\eta) \, \label{tt-constraint}\\
 t_\mu x^\mu &= 0 \ \label{x-t-orth}
\end{align}
\end{subequations}
which arise from the special properties of $\cH_n$. 
Here  $x^\mu:\, \cM^{3,1}\hookrightarrow \R^{3,1}$ is interpreted as Cartesian  
coordinate functions, and $\eta$ plays the role of a time parameter, defined via
\begin{align}
 x^4 = R \sinh(\eta) \ .
 \label{x4-eta-def}
\end{align}
Hence $\eta = const$ defines a foliation of $\cM^{3,1}$ into space-like surfaces $H^3$; this 
will be related to the scale parameter of a FLRW cosmology \eqref{a-eta} with $k=-1$.
Note that
%%H  ``the sign of'' added
the sign of $\eta$ distinguishes the two degenerate sheets of $\cM^{3,1}$, 
cf. figure \ref{fig:projection}. 
The $t^\mu$ generators clearly describe the $S^2$ fiber over $\cM^{3,1}$, which is 
 space-like due to \eq{x-t-orth}. 
These generators satisfy the Poisson brackets
\begin{align}
 \{x^\mu,x^\nu\} &= \theta^{\mu\nu}  = - r^2 R^2 \{t^\mu,t^\nu\}  ,  \nn\\
 \{t^\mu,x^\nu\} &= \frac{x^4}{R} \eta^{\mu\nu} \label{Poisson-brackets} \ ,
\end{align}
and the Poisson tensor $\theta^{\mu\nu}$ satisfies the constraints
\begin{subequations}
\label{geometry-H-theta}
\begin{align}
 t_\mu \theta^{\mu\a} &= - \sinh(\eta) x^\a , \\
 x_\mu \theta^{\mu\a} &= - r^2 R^2 \sinh(\eta) t^\a , \label{x-theta-contract}\\
 \eta_{\mu\nu}\theta^{\mu\a} \theta^{\nu\b} &= R^2 r^2 \eta^{\a\b} - R^2 r^4 
t^\a t^\b + r^2 x^\a x^\b  
% 
% \theta^{\mu\nu}\theta_{\mu\nu} &= 2R^2r^2\big(2-\cosh^2(\eta)\big)
\end{align}
\end{subequations}
Due to the self-duality relations of $\theta^{ab}$ on $H^4$,
$\theta^{\mu\nu}$ can be expressed in terms of $t^\mu$ as  \cite{Sperling:2018xrm}
\begin{align}
 \theta^{\mu\nu} &= \frac{r^2}{\cosh^2(\eta)} 
   \Big(\sinh(\eta) (x^\mu t^\nu - x^\nu t^\mu) +  \epsilon^{\mu\nu\a\b} x_\a t_\b \Big) \ ,
 \label{theta-P-relation} \, 
\end{align}
The above commutation relations imply
\begin{align}
 \{t_\mu,\phi\} =   \sinh(\eta)\del_\mu \phi 
 \label{del-t-rel}
\end{align}
for $\phi = \phi(x)$, which
suggest that $T^\mu \sim t^\mu$ can be viewed as momentum generators on $\cM^{3,1}$.

\subsection{Algebra of functions and higher-spin}
\label{sec:higher-spin-alg}

The full algebra of functions $\End(\cH_n) = \oplus\, \cC^s$ still decomposes into sectors $\cC^s$  \eq{EndH-Cs-decomposition}
corresponding to spin $s$ harmonics on the $S^2$ fiber, which is respected by  
the d'Alembertian on $\cM^{3,1}$ because
 \begin{align}
 [\cS^2,\Box] = 0    \ .
\end{align}
This decomposition is hence compatible with the kinematics defined through $\Box$.
We can write the most general $\phi\in \cC^s\subset \End(\cH)$ as a function 
$\phi = \phi(x,t)$ on $\C P^{1,2}$, which is identified with a totally symmetric traceless space-like 
rank $s$ tensor fields on $\cM^{3,1}$ via
\begin{align}
 \phi^{(s)} = \phi_{\mu_1 ... \mu_s}(x) t^{\mu_1} ... t^{\mu_s} \ ,
 \qquad \phi_{\mu_1 ... \mu_s} x^{\mu_i} = 0\ .
 \label{Cs-explicit}
\end{align}
The restriction to space-like and traceless tensors follows from the constraints
\eqref{geometry-H-M}. However this  just amounts to a gauge,
and the physical tensor fields (e.g. the metric perturbation $h_{\mu\nu}$ \eq{gravitons-H1}, \eq{Hmunu-tr})
can arise from $\phi^{(s)}$ in a different way, via
\begin{align}
  \tilde\phi_{\mu_1 \ldots  \mu_s}(x) := [\{x_{\mu_1},\ldots\{x_{\mu_s},\phi^{(s)}\}\ldots\}]_0 \ .
 \label{psi-iso-2}
\end{align}
Here $[]_0$ denotes the projection to $\cC^0$. These are totally symmetric tensor fields which are no longer space-like 
but satisfy other constraints.

To demonstrate consistency of the theory, we must show that the time-like components in tensor fields 
do not lead to negative-norm states, i.e.  ghosts.
Basically the only known way to define such a consistent quantum theory is via gauge theory, where the unphysical 
negative norm modes can be removed while effectively preserving Lorentz invariance. 
The prime examples are Yang-Mills gauge theory for spin 1 and diffeomorphism-invariant
general relativity for spin 2. For higher spin this is achieved by Vasiliev's higher spin theory 
\cite{Vasiliev:1990en,Didenko:2014dwa}, however no action formulation is known.
%In particular, one can choose a gauge where the  fields have no time-like components, as in \eq{Cs-explicit}.

The present framework provides a slightly different solution to this problem. A key ingredient is 
the space-like constraint \eq{Cs-explicit}, which
is built in a priori\footnote{This is not the case for the tangential fluctuation modes $\cA^\mu$,
which do have time-like components a priori. However these are then removed as in Yang-Mills theory, as 
explained in section \ref{sec:physical}.}. This comes of course at the expense of manifest local 
Lorentz invariance, and we will see that only the space-like isometries $SO(3,1)$ are manifest,
while boosts are not. 
%%V3: refurmulation
However, the model appears to have sufficient extra symmetries and properties that protect the theory 
from significant Lorentz violations. In particular,
we will see  that all modes propagate in the  standard relativistic way.
Moreover there is a large 
higher-spin gauge invariance, which includes diffeomorphisms that preserve a  
volume form, albeit acting in a  non-standard way via the NC structure. 
These topics will be discussed in more detail below.

\paragraph{Higher-spin gauge transformations.}

The origin of the higher-spin gauge symmetry can already be understood at this stage:
They arise from gauge transformations \eq{gaugetrafo-matrix} $Y^a \to U Y^a U^{-1}$, or infinitesimally
for $U = e^{i\L}$
\begin{align}
 Y^a \to i[\L,Y^a], \qquad \L = \L^\dagger \in \End(\cH) \ .
\end{align}
We can parametrize the generator $\L\in \cC^s \subset End(\cH)$ in two equivalent ways
\begin{align}
 \L &= \L_{\mu_1 ... \mu_s}(x) t^{\mu_1} ... t^{\mu_s}   
 \ = \ \tilde\L_{\mu_1...\mu_s;\nu_1...\nu_s}(x) \theta^{\mu_1\nu_1} ... \theta^{\mu_s\nu_s} \qquad \in \cC^s \ .
\end{align}
The first form is reminiscent of a frame-like higher spin generator identifying $t^\mu$ as momentum generator \eq{del-t-rel}, 
while the second form can be viewed as a gauge transformation taking values in
$U(\mso(3,1))$ corresponding to Young diagrams 
${\tiny \Yvcentermath1 \young(\mu\mu\mu,\nu\nu\nu) }$, as
$\theta^{\mu\nu}$ are generators of $\mso(3,1)$ \eq{M-M-Poisson}.
Both are equivalent due to the implicit constraints.
Using the first form, the spin 1 gauge transformations
\begin{align}
 \L = \L_\mu(x) t^\mu 
 \label{lambda-spin-1}
\end{align}
are identified with space-like vector fields $\L_\mu(x)$. Even though $\L_\mu$ is space-like, the 
resulting gauge transformation includes time-like directions, because it acts via the commutator or the Poisson bracket.
This also leads to a mixing of the different spin sectors, and
the specific transformation depends on the object under consideration. 
For example, functions $\phi\in\cC^0$ transform as 
\begin{align}
 \d_\L \phi &= \{\L,\phi\} 
   =  \xi^\mu  \del_\mu\phi , \qquad 
   \xi^\mu = \{\L,x^\mu\} \ .
%    &= \frac{1}{3} \left(\sinh(\eta)\left( 3 \L^\mu \del_\mu + ({\rm div} \L) \t 
%    - (\t \L^\mu) \del_\mu \right)\phi
%     + x_\g \varepsilon^{\g\mu\a\b}\del_\a \L_\mu \del_\b \phi\right) 
%     \ \ (+ \cC^2)
\label{gaugetrafo-Lambda}
\end{align}
This looks like a diffeomorphism generated by a vector field $\xi^\mu$, however 
$\xi^\mu$ is a higher-spin-valued vector field here, i.e. 
\begin{align}
 \xi^\mu &= \xi^\mu_{(0)} \ + \xi^\mu_{(2)} \qquad \in \cC^0 \ \oplus \ \cC^2 \ .
 \label{xi-general-spin1}
\end{align}
We can understand this by decomposing
 $\L_\mu$ \eq{lambda-spin-1} into divergence-free and pure divergence part,
\begin{align}
 \L = \L^{(1,0)} + \L^{(1,1)}, \qquad 
 \del^\mu\L_\mu^{(1,0)} = 0, \quad \L^{(1,1)}_\mu = \del_\mu\L(x) \ 
 \label{diffeo-decomp}
\end{align}
anticipating the notation in \eq{primal-descendant}.
Then $\xi_\mu$ decomposes  accordingly
into the space-like divergence-free field (see section 9.2 in \cite{Steinacker:2019awe})
\begin{align}
 \xi_\mu[\L^{(1,0)}] &= - \frac{1}{3} \sinh(\eta) (\t+3)\L^{(1,0)}_\mu   
  + \frac{1}{3} x_\b \varepsilon^{\b \mu \a\nu} \del_\nu\L^{(1,0)}_\a , \nn\\
  \del^\mu \xi_\mu  &= 0 =  x^\mu \xi_\mu
  \label{spacelike-A10}
\end{align}
(hence in radiation gauge with 2 d.o.f.), and a scalar mode
\begin{align}
 \xi_\mu[\L^{(1,1)}] &=\frac{r^2R}{3}
 \sinh(\eta) (x_\mu\del^\a\del_\a\L  - (\t+3)\del_\mu\L)  , \nn\\
  \del^\mu\xi_\mu  &= - \frac{1}{R^2\sinh^2(\eta)} x^\mu \xi_\mu 
  \label{A-Dphi0-explicit}
\end{align}
which is neither space-like nor divergence-free. 
The constraint can be written covariantly as
\begin{align}
  \nabla_\mu (\b^{3} \xi^\mu) = 0 \ ,
 \label{gauge-diffeo-constraint}
\end{align}
where $\nabla$ is the covariant derivative associated to the effective metric \eq{eff-metric-G}.
Therefore $\xi^\mu$ corresponds to a volume-preserving diffeomorphism, up to the factor $\b^{3}$.
The component
$\xi^\mu_{(2)} = -\del_\nu\phi_{\a} [\theta^{\mu\nu}t^{\a}]_2 \in \cC^2$ in \eq{xi-general-spin1} 
is some derivative contribution which we 
will not pursue any further. 
To put this into context, we note
that these spin 1 gauge transformations will indeed induce the usual
transformations of graviton modes \eq{puregauge-diffeo-rel}
\begin{align}
  \d_\L G^{\mu\nu} 
    &=  \nabla^\mu\xi^\nu + \nabla^\nu \xi^\mu \ .
 \label{puregauge-diffeo-rel-2}
\end{align}
%\paragraph{Volume-preserving $\hs$ diffeomorphisms and normal frame.}
Hence the spin 1 gauge transformations can be interpreted as 
volume-preserving diffeomorphisms in the sense \eq{gauge-diffeo-constraint}.
More generally,  \eq{gaugetrafo-Lambda} provides a nice geometrical interpretation of 
higher-spin gauge transformations: they are simply  
Hamiltonian vector fields on $\C P^{1,2}$ which generate the symplectomorphism group.
Since these preserve the symplectic volume on $\C P^{1,2}$, the resulting diffeomorphisms on $\cM^{3,1}$
are volume-preserving in the sense of \eq{gauge-diffeo-constraint}.
%and encode only 3 rather than 4 degrees of freedom (dof).

\paragraph{$SO(3,1)$ substructure.}

The decomposition of diffeomorphisms found in \eq{diffeo-decomp} 
illustrates the sub-structure of tensor fields resulting from the reduced $SO(3,1)$ covariance,
indicated by the superscripts of $\L$. 
All ``fundamental'' tensor fields \eq{Cs-explicit} are space-like, and decompose further into divergence-free and 
pure divergence modes.
There is a systematic organization  developed in \cite{Sperling:2019xar}, which is based on 
the underlying $\mso(4,2)$ structure. Consider the $SO(3,1)$ -invariant derivation
\begin{align}
 D\phi &:= \{x^4,\phi\} \ 
  = r^2 R^2 \frac{1}{x^4} t^\mu \{t_\mu,\phi\}   
  = -\frac{1}{x^4}x_\mu\{x^\mu,\phi\} \nn\\
   &= r^2 R\, t^{\mu_1}\ldots t^{\mu_s} t^\mu \, \nabla^{(3)}_\mu\phi_{\mu_1\ldots\mu_s}(x) 
 \label{D-properties}
\end{align}
where  $\nabla^{(3)}$ is the covariant derivative along the 
space-like $H^3 \subset \cM^{3,1}$.
Hence $D$ relates the different spin sectors in \eqref{EndH-Cs-decomposition}:
\begin{align}
  D = D^- + D^+: \ \cC^{s} \ &\to \cC^{s-1} \oplus \cC^{s+1}, \qquad
    D^\pm \phi^{(s)} = [D\phi^{(s)}]_{s\pm 1} \ 
    \label{D-properties-2} 
\end{align}
where $[.]_{s}$ denotes the projection to $\cC^{s}$ defined through 
\eqref{EndH-Cs-decomposition}.
It is easy to see that  $(D^+)^\dagger = - D^-$
w.r.t. the canonical invariant inner product.
% and any $\phi\in\cC^{s}$ can be written as  $D^-() + D^+()$.
% In particular, $D^+\cC^{s}$ is surjective except for $\cC^{(s,0)}$.
Explicitly, $D x^\mu = r^2 R\, t^\mu$ and $D t^\mu =  R^{-1}\, x^\mu$. 
% Then 
% \begin{align}
%   D^\pm: \quad \cC^{(s,k)} &\to \cC^{(s-1,k-1)}  \ .
%   \label{D-refined}
% \end{align}
In particular,  $\cC^{(s,0)} \subset \cC^{s}$ is the space of divergence-free traceless
space-like rank $s$ tensor fields on $\cM^{3,1}$, in radiation gauge.
We can then organize the $\cC^{s}$ modes
into primals and descendants \cite{Sperling:2018xrm,Steinacker:2019awe}
\begin{align}
 \cC^{(s,0)} &= \{\phi\in\cC^s; \ D^- \phi = 0 \} &  \mbox{... primal fields}\ \nn\\
 \cC^{(s+k,k)} &= (D^+)^k\cC^{(s,0)} &  \mbox{... descendants} \ \
\label{primal-descendant}
\end{align}
The $D^+$ and $D^-$ satisfy   ladder-type commutation relations
\begin{align}
\ \  D^+ D^- \phi^{(s,k)} = \big(a_{s,k} D^- D^+ + b_{s,k} \big) \phi^{(s,k)},
\qquad  \phi^{(s,k)} = (D^+)^k \phi^{(s-k,0)} \ 
  \label{D+D--CR-general}
\end{align}
with constants $a_{s,k}, b_{s,k}$ given  in \cite{Steinacker:2019awe}.
This can be shown using the $\mso(4,2)$ structure, as well as the special properties 
of the minireps $\cH_n$.
This sub-structure encodes two different concepts of spin on the FRW background, which 
arise from the space-like foliation:
$\cS^2=2s(s+1)$ measures the 4-dimensional spin on $H^4$,
while $s-k$ measures the 3-dimensional spin of $\cC^{(s,k)}$ on $H^3$. 
%Nevertheless, local Lorentz invariance should be largely restored through  gauge invariance.

\paragraph{Effective metric and d'Alembertian.}

In the matrix model framework, the effective metric on any given background 
is obtained by rewriting the
kinetic term in covariant form \cite{Sperling:2019xar,Steinacker:2010rh}.
Consider e.g. a transversal fluctuation $\phi = Y^4$ in the model \eq{MM-Mink}, on 
the $\cM^{3,1}$ background $Y^\mu=T^\mu$ under consideration. It suffices to consider scalar fluctuations $\phi = \phi(x)$.
Then the action for $\phi$ is
\begin{align}
S[\phi] =  - \Tr [T^\mu,\phi][T_\mu,\phi] 
\sim \int d^4 x\,\sqrt{|G|}G^{\mu\nu}\del_\mu\phi \del_\nu \phi \ 
\label{scalar-action-metric}
\end{align}
where
\cite{Sperling:2019xar}
\begin{align}
   G^{\mu\nu} &= \sinh^{-3}(\eta)\, \g^{\mu\nu},
   \qquad  \g^{\a\b} = \eta_{\mu\nu}\theta^{\mu\a}\theta^{\nu\b} 
  = \sinh^2(\eta) \eta^{\a\b} \ 
   \label{eff-metric-G}
\end{align}
up to an irrelevant constant.
This  is a 
$SO(3,1)$-invariant FLRW metric with signature $(-+++)$, 
\begin{align}
 d s^2_G = G_{\mu\nu} d x^\mu d x^\nu 
   &= -R^2 \sinh^3(\eta) d \eta^2 + R^2\sinh(\eta) \cosh^2(\eta)\,d \Sigma^2 \ \nn\\
   &= -d t^2 + a^2(t)d\Sigma^2 \, .
   \label{eff-metric-FRW}
\end{align}
where $d\Sigma^2 =  d\chi^2 + \sinh^2(\chi)d\Omega^2$ is the metric on the 
unit hyperboloid $H^3$.
We can read off the cosmic scale parameter $a(t)$  
\begin{align}
a(t)^2 &=  R^2\sinh(\eta) \cosh^2(\eta) \ \stackrel {t\to\infty}{\sim}  \  R^2\sinh^3(\eta) ,  \label{a-eta}\\
d t &=  R \sinh(\eta)^{\frac{3}{2}} d\eta \ 
\end{align}
which leads to $a(t) \sim \frac 32 t$ for late times\footnote{Another FLRW solution of the IKKT model with  similar structure to the present one  but $k=+1$  
was discussed in \cite{Steinacker:2017vqw}, however it leads to a cosmological evolution which is less realistic 
than a coasting evolution $a(t) \propto t$, cf.  \cite{Steinacker:2017bhb}.}, and $a(t) \sim t^{1/5}$ near the Big Bounce.
This metric can also be extracted from the % $SO(3,1)$-invariant 
matrix d'Alembertian \eq{dAlembert-fuzzy-def} 
\begin{align}
 \Box := [T^\mu,[T_\mu,.]] \  \sim \ -\{t^\mu,\{t_\mu,.\}\}  \ 
   = \sinh^{3}(\eta) \Box_G
 \label{Box-def}
\end{align}
acting on  $\phi \in \cC^0$, 
where $\Box_G = -\frac{1}{\sqrt{|G|}}\del_\mu\big(\sqrt{|G|}\, G^{\mu\nu}\del_\nu\big)$.

\paragraph{Remarks on local Lorentz (non-)invariance.}

The local isometry group of the background $\cM^{3,1}$  at any point comprises only space-like 
$SO(3)$ rotations, which is part of the global symmetry of the matrix model.
This reduction  reflects the presence of a global time-like vector field 
$\t = x^\mu \del_\mu$ of the FLRW geometry, as in ordinary GR,
which is of course not a problem.
The important question is if the local physics is invariant under local Lorentz transformations,
up to corrections which enter only at the IR scale set by the background.

To put this into perspective, we recall the case of 
noncommutative field theory on 
the Moyal-Weyl quantum plane \cite{Douglas:2001ba,Szabo:2001kg}, which carries an explicit Poisson tensor 
$\theta^{\mu\nu} = \{x^\mu,x^\nu\}$.
This tensor explicitly breaks Lorentz-invariance, with scale set by the UV scale of noncommutativity.
This leads to significant Lorentz violations, notably in the quantum effects due to loop contributions, 
which probe the UV regime of the theory.

In the present setting, the violation of local Lorentz invariance is expected to be much milder.
The Poisson tensor $\theta^{\mu\nu}$ disappears on $\cM^{3,1}$
upon averaging over the internal fiber (reflecting the  local $SO(3)$ invariance), which leaves only the  
contributions from the cosmic vector field $\t$. 
On more general backgrounds, 
$\t$ is subsumed by a torsion tensor $\tensor{T}{_\mu_\nu^\r}$ \cite{Steinacker:2020xph}, 
and both are clearly geometric quantities defining an IR scale.
This suggests that the Lorentz violation is weak in the UV and can be attributed to 
some geometric quantity, which may incorporate  new physics.
Finally, the volume-preserving diffeomorphisms \eq{gaugetrafo-Lambda} should
allow to go to weaker local normal coordinates, such that the metric at any given point takes the 
canonical form $G_{\mu\nu} = c(x) \eta_{\mu\nu}$ up to a conformal factor. 
Thus one can ``almost'' go to local free-falling elevator frames.
This also suggests the 
presence of extra physical degrees of freedom in the metric, which will be discussed further
in section \ref{sec:graviton}.
However, all these issues need to be studied in more detail in future work.

One might contemplate the idea of preserving manifest local 
Lorentz invariance in an extended model where $\End(\cH)$
describes some local geometry $\R^{3,1} \times\cK$. 
This includes many attractive examples such as fuzzy de Sitter spaces
and others \cite{Buric:2017yes,Gazeau:2009mi}, see also \cite{Doplicher:1994tu}.
However, then $\cK$ must be a homogeneous space of $SO(3,1)$, which is necessarily non-compact.
Then an expansion in discrete harmonics on $\cK$ 
no longer makes sense, and the theory would display a higher-dimensional behavior.
Thus it seems quite hopeless to obtain a consistent 4-dimensional theory
from the fluctuations.
This is avoided in the the present  model, at the price of only partially manifest 
local Lorentz invariance.
In any case, the issue of (local) Lorentz invariance needs further clarification.

%%%%%%%%%%%%%%%%%%%%%%%%%%%%%%%%%%%%%%%%%%%%%%%%%%%
%%%%%%%%%%%%%%%%%%%%%%%%%%%%%%%%%%%%%%%%%%%%%%%%%%%
%
\subsection{Matrix model fluctuations and higher-spin Yang-Mills theory}
\label{sec:fluctuations}

Now we return to the  noncommutative setting, and 
define a dynamical model for the fuzzy $\cM^{3,1}$ space-time.
Consider again the Yang-Mills matrix model \eq{MM-Mink} with specific mass term,
\begin{align}
 S[Y] &= \frac 1{g^2}\Tr \Big([Y^\mu,Y^\nu][Y_{\mu},Y_{\nu}] \, 
  +\frac{6}{R^2} Y^\mu Y_\mu  \Big) \ . 
 \label{bosonic-action}
\end{align}
As observed in \cite{Sperling:2019xar}, $\cM^{3,1}$ is indeed a solution 
of this model\footnote{This  ''momentum'' embedding via $T^\mu$ has some similarity with 
the ideas in \cite{Hanada:2005vr} but avoids excessive dof and the associated ghost issues, 
cf. \cite{Sakai:2019cmj}.
The positive mass parameter in \eqref{bosonic-action} simply sets the scale of the background.
For negative mass parameter, $X^\mu$ would be a solution \cite{Steinacker:2017bhb}, 
but the fluctuation analysis would be less clear.}, through
%V3 added as paragraph
\begin{align}
 Y^\mu = T^\mu
 \label{background-solution-M}
\end{align}
due to \eq{Box-X-T}.
Now consider  tangential
deformations of this background solution, i.e.
\begin{align}
 Y^\mu = T^\mu  + \cA^\mu \ , 
\end{align}
where $\cA^\mu \in \End(\cH_n) \otimes \R^4$ is an arbitrary Hermitian fluctuation.
The  Yang-Mills action \eqref{bosonic-action}  can be expanded around the solution as
\begin{align}
 S[Y] = S[T]  +  S_2[\cA] + O(\cA^3) \ ,  
 \end{align}
 and the quadratic fluctuations are  governed by  
 \begin{align}
S_2[\cA] = -\frac{2}{g^2} \,\Tr \left( \cA_\mu 
\Big(\cD^2 -\frac{3}{R^2}\Big) \cA^\mu + \cG\left(\cA\right)^2 \right) .
\label{eff-S-expand}
\end{align}
This involves the vector d'Alembertian on $\cM^{3,1}$
\begin{align}
\cD^2 \cA =  \left(\Box  - 2\cI \right)\cA  
\label{vector-Laplacian}
\end{align}
(cf. \eq{Box-def})
which is an $SO(3,1)$ intertwiner, as well as 
\begin{align}
 \cI (\cA)^\mu := - [[ Y^\mu, Y^\nu],\cA_\nu] =  \frac{i}{r^2 R^2} 
[\Theta^{\mu\nu},\cA_\nu] 
 =: -\frac{1}{r^2 R^2}\tilde\cI (\cA)^\mu \ 
 \label{tilde-I-NC}
\end{align}
using \eqref{XT-CR}.
As usual in Yang-Mills theories, $\cA$ transforms under gauge transformations as
\begin{align}
 \d_\L\cA = -i[T^\mu  + \cA^\mu,\L] \sim \{t^\mu,\L\}  + \{\cA^\mu,\L\}
\end{align}
for any $\L\in\cC$,
and the scalar ghost mode
 \begin{align}
\cG(\cA) = -i [T^\mu,\cA_\mu] \sim \{t^\mu,\cA_\mu\}   
 \label{gaugefix-intertwiner}
 \end{align}
should be removed.
This is achieved by adding a gauge-fixing term $-\cG(\cA)^2$ to the action
as well as the corresponding Faddeev-Popov (or BRST) ghost. Then the quadratic 
action becomes 
\begin{align}
 S_2[\cA] + S_{g.f} + S_{ghost} &= -\frac{2}{g^2}\Tr\, 
\left( \cA_\mu \Big(\cD^2  -\frac{3}{R^2} \Big) \cA^\mu + 2 \obar{c} \Box 
c \right) \ 
\label{eff-S-gaugefixed}
\end{align}
where $c$ denotes the  BRST ghost; see e.g.\ \cite{Blaschke:2011qu} 
for more details.
% 

%%%%%%%%%%%%%%%%%%%%%%%%%%%%%%%%%%%%%%%%%%%%% 
% 
\subsection{Fluctuation modes}
\label{sec:modes}

We now expand the vector modes into higher spin modes according to 
\eqref{EndH-Cs-decomposition}, \eqref{Cs-explicit}
\begin{align}
 \cA^\mu &= A^{\mu}(x) + A^{\mu}_\a(x)\, t^\a +   A^{\mu}_{\a\b}(x)\, t^\a t^\b + \ldots 
  \ \in \ \cC^0   \oplus  \cC^1  \oplus  \cC^2  \oplus \ \ldots
 \label{A-M31-spins}
\end{align}
We need to find explicitly all eigenmodes of $\cD^2$. This can be achieved using the $\mso(4,2)$
structure and suitable intertwiners.
The result is as follows \cite{Steinacker:2019awe}:
First, for any given  $\phi^{(s)} \in \cC^s$ we define the  fluctuation modes
\begin{align}
\label{A2-mode-ansatz}
 \cA_\mu^{(g)}[\phi^{(s)}] &= \{t_\mu,\phi^{(s)}\}  \quad \in \cC^{s}\,,
\\
 \cA_\mu^{(+)}[\phi^{(s)}] &= \{x_\mu,\phi^{(s)}\}\big|_{s+1} \ \equiv  
\{x_\mu,\phi^{(s)}\}_+ \quad \in \cC^{s+1} \,, \\
 \cA_\mu^{(-)}[\phi^{(s)}] &= \{x_\mu,\phi^{(s)}\}\big|_{s-1} \  \equiv  
\{x_\mu,\phi^{(s)}\}_- \quad \in \cC^{s-1} \nn\\
 \cA_\mu^{(n)}[\phi^{(s)}] &= D^+\cA_\mu^{(-)}[\phi^{(s)}] \qquad \in \cC^{s} \ . \
\end{align}
Then for any eigenmode of $\Box\phi^{(s)} = m^2 \phi^{(s)}$ 
we obtain 4-tuples of {\em ''regular`` eigenmodes} $\tilde\cA_\mu^{(i)}[\phi^{(s)}] \in \cC^{s}\otimes \R^4$  of $\cD^2$ 
\begin{align}
 \tilde\cA^{(i)}[\phi] = \begin{pmatrix}
      \cA^{(+)}[D^-\phi] \\ \cA^{(-)}[D^+\phi] \\ \cA^{(n)}[\phi] \\ r^2 R \cA^{(g)}[\phi]
    \end{pmatrix} , \qquad i,j\in\{+,-,n,g\} \ 
    \label{A-tilde-def}
\end{align}
for $\phi = \phi^{(s)}$ dropping the index $\mu$,
with the same eigenvalue 
\begin{align}
 \boxed{ \
\begin{aligned}
 \cD^2 \tilde\cA^{(+)}[\phi] &= \big(m^2 + \frac{3}{R^2}\big) \tilde\cA^{(+)}[\phi]    \\
 \cD^2 \tilde\cA^{(-)}[\phi] &= \big(m^2 + \frac{3}{R^2}\big) \,\tilde\cA^{(-)}[\phi] \\
 \cD^2 \tilde\cA^{(g)}[\phi] &= \big(m^2 + \frac{3}{R^2}\big) \, \tilde\cA^{(g)}[\phi]  \\
 \cD^2 \tilde\cA^{(n)}[\phi] &= \big(m^2 + \frac{3}{R^2}\big) \, \tilde\cA^{(n)}[\phi] \ . \
\end{aligned}
 }
  \label{Apmg-degeneracy}
\end{align}
%Linear independence of these modes will be established in section \ref{sec:inner}. 
There is  one ''special`` mode which is not covered by the regular $\tilde\cA^{(i)}$, namely 
$\cA^{(-)}[\phi^{(s,0)}]$ with
\begin{align}
 \cD^2 \cA^{(-)}[\phi^{(s,0)}] 
 = \big(\Box + \frac{-2s+3}{R^2}\big)\cA^{(-)}[\phi^{(s,0)}] \ . 
\label{special-s0}
\end{align}
We will see that it is orthogonal to all regular modes, and altogether these modes are complete.
Hence diagonalizing $\cD^2$ is reduced  to diagonalizing $\Box$ on $\cC^s$. 
In particular,
we obtain the following on-shell modes $\big(\cD^2  -\frac{3}{R^2} \big) \cA = 0$ 
\begin{align}
\{\tilde\cA^{(+)}[\phi^{(s)}], \tilde\cA^{(-)}[\phi^{(s)}], 
  \tilde\cA^{(g)}[\phi^{(s)}],\tilde\cA^{(n)}[\phi^{(s)}]\}  \qquad  &\text{for } 
       \quad   \ \Box \phi^{(s)} = 0  \ \nn\\
\cA^{(-)}[\phi^{(s,0)}]  \qquad  &\text{for } 
       \quad  \ \big(\Box - \frac{2s}{R^2} \big)\phi^{(s,0)} = 0 \ .
  \label{on-shell-all}
\end{align}
These modes are not orthogonal yet, but it is possible to find a
basis of orthogonal eigenmodes by diagonalizing the intertwiner $\cI$ \eq{tilde-I-NC}, 
which commutes with $\cD^2$
\begin{align}
 [\cI,\cD^2] = 0 \ .
\end{align}
This turns out to be rather tedious, and requires the sub-structure of 
$\phi^{(s,k)}$ modes as defined in \eq{primal-descendant}. 
The relations \eq{D+D--CR-general} then allows to compute all the 
inner products and eigenvalues explicitly, which is carried out in  \cite{Steinacker:2019awe}.
It turns out that $\tilde\cA^{(n)}[\phi^{(s,s)}]$ is redundant, and 
$\tilde\cA^{(+)}[\phi^{(s,0)}]\equiv 0$.
One can then show the following completeness statement:
\begin{thm}
The  $\tilde\cA^{(i)}[\phi^{(s)}]$ modes \eq{A-tilde-def} along with the $\cA^{(-)}[\phi^{(s,0)}]$ for all $s\geq 0$
span the space of all fluctuations $\cA$.
A basis is  obtained by dropping $\tilde\cA^{(n)}[\phi^{(s,s)}]$ and  $\tilde\cA^{(+)}[\phi^{(s,0)}]$.
\label{thm:basis}
\end{thm}
This completes the classification of off-shell modes.

\subsection{Physical constraint, Hilbert space and no-ghost theorem}
\label{sec:physical}

Now consider the on-shell modes.
We first observe that an (admissible, i.e. square-integrable) fluctuation mode $\cA$
satisfies the gauge-fixing condition 
$\{t^\mu,\cA_\mu\} = 0$ if and only if it is orthogonal to all pure gauge modes,
\begin{align}
 \langle \cA^{(g)},\cA\rangle \equiv \int \cA^{(g)\mu}\cA_\mu = 0 \ .
 \label{gaugefix-inner}
\end{align}
where the $SO(4,2)$ - invariant integral arises from the trace on $\End(\cH_n)$.
Now consider an on-shell mode $\cA\in\cC^s$ in some 4-dimensional mode space $\tilde \cA^{(i)}[\phi], \ i\in\{+-ng\}$ determined by some $\phi\in\cC^{(s,k)}$
with $\Box\phi = 0$ and  $s>k>0$. 
One can show that this 4-dimensional space of modes
has signature $(+++-)$ and $\cA^{(g)}$ is null. Then
the gauge-fixing constraint \eq{gaugefix-inner} leads to a 3-dimensional subspace 
with signature $(++0)$, which contains $\cA^{(g)}$.
Then the usual definition 
\begin{align}
 \cH_{\rm phys} = \{\mbox{gauge-fixed on-shell modes}\}/_{\{\mbox{pure gauge modes}\}  }
 \label{H-phys}
\end{align}
leads  to 2 modes with positive norm. 
By orthogonalizing all the 
eigenmodes \eq{Apmg-degeneracy} of $\cD^2$,
one can similarly establish the no-ghost theorem \cite{Steinacker:2019awe}
\begin{thm}
 The space  $\cH_{\rm phys}$ \eq{H-phys} of admissible solutions of 
 $\big(\cD^2-\frac{3}{R^2}\big)\cA = 0$ which are gauge-fixed $\{t^\mu,\cA_\mu\} = 0$ 
 modulo pure gauge modes inherits a positive-definite inner product, and forms a Hilbert space.
\label{thm:no-ghost}
\end{thm}
''Admissible`` means that the modes are square-integrable on $H^4$ 
or equivalently on $\cM^{3,1}$, more precisely that they
live in principal series unitary representations of $SO(4,1)$. 
On-shell, this is essentially equivalent to the requirement 
that they are square-integrable on space-like slices $H^3$.
Indeed, the on-shell relation $\Box\phi = 0$ 
determines $C^2[SO(4,1)]$ via \eq{dAlembert-fuzzy-def} from $C^2[SO(3,1)]$, 
correspond to an irreducible 
tensor field on the space-like $H^3$.
In other words, the state at any given time-slice $H^3$ 
completely determines the time evolution (up to time direction).
Hence one obtains the standard picture of time evolution even though time does not commute, and
the time evolution is completely captured by $SO(4,1)$ group theory, even though 
$\cM^{3,1}$ admits only space-like  $SO(3,1)$ isometries. 
%In other words, the group theory is even more powerful than  expected.

Explicitly,
this gives the following physical modes:

\paragraph{The physical modes $\cA_\mu \in \cC^0$.}

The off-shell modes $\cA_\mu \in \cC^0$ comprise 
the spin 1 mode $\cA^{(-)}[\phi^{(1)}]$ and 
the spin 0 modes $\cA^{(g)}[\phi^{(0)}]$ are in $\cC^0 \otimes \R^4$.
%These modes are elaborated explicitly in section \ref{sec:C0-modes-explicit}.
Among these, only the spin 1 modes $\cA^{(-)}[\phi^{(1,0)}]$ are physical, and 
\begin{align}
  \cH_{\rm phys} \cap \cC^0 = \{\cA^{(-)}[\phi] \ \mbox{for} \  \phi\in\cC^{(1,0)},\ 
   \big(\Box - \frac{2}{R^2} \big)\phi = 0 \} \ .
\end{align}
These modes satisfy $\del^\mu\cA_\mu = 0 = x^\mu \cA_\mu$,
and describe a spin 1 Yang-Mills (or Maxwell) field.

\paragraph{The physical modes $\cA_\mu\in \cC^1$.}

This space comprises 12 
off-shell modes $\cA^{(-)}[\phi^{(2)}]$, $\cA^{(n)}[\phi^{(1)}]$, $\cA^{(g)}[\phi^{(1)}]$ 
and $\cA^{(+)}[\phi^{(0)}]$. 
Among these, all  $\cA^{(-)}[\phi^{2)}]$ are physical, 
and  there are no further physical states in this sector:
\begin{align}
  \cH_{\rm phys} \cap \cC^1 = \{\cA^{(-)}[\phi]\ \mbox{for} \  \phi\in\cC^{(2,*)}, \ 
   \big(\Box - \frac{4}{R^2} \big)\phi = 0 \} \ .
   \label{Hphys-C1}
\end{align}
They satisfy $\{t^\mu,\cA_\mu\} = 0$, and $x^\mu \cA_\mu[\phi^{(2,0)}]=0$.
These modes govern the linearized gravity sector,
as discussed below.

\paragraph{Generic physical modes $\cA_\mu\in \cC^s$ with $s\geq 2$.}

Finally in the generic case $s\geq 2$, the physical constraint $\{t^\mu,\cA_\mu\} = 0$  
must be solved directly. This leads to the following modes:
There is one  physical mode determined by $\phi^{(s,0)}$, 
which we can choose to be a linear combination  
\begin{align}
 \{c_-\tilde\cA^{(-)}[\phi^{(s,0)}] + \tilde\cA^{(n)}[\phi^{(s,0)}] \ \ \mbox{for}  \ 
   \Box\phi^{(s,0)} = 0 \ \} \ \subset \ \cH_{\rm phys} \cap \cC^s  
   \label{Hphys-Cs-primal}
\end{align}
where $c_-$ is determined by solving the above constraint.
Next, there is  one physical scalar mode  determined by $\phi^{(s,s)}$
for each $s\neq 0$, which we can choose to be 
\begin{align}
 \{\tilde\cA^{(-)}[\phi^{(s,s)}] + c_+\tilde\cA^{(+)}[\phi^{(s,s)}] \ \ \mbox{for}  \ 
   \Box\phi^{(s,s)} = 0 \ \} \ \subset \ \cH_{\rm phys} \cap \cC^s  \ .
   \label{Hphys-Cs-scalar}
\end{align}
Finally for  $s\neq k\neq 0 $, there are two physical modes  determined by $\phi^{(s,k)}$, 
which we can choose to be 
\begin{align}
 \{\tilde\cA^{(-)}[\phi^{(s,k)}] + c_+\tilde\cA^{(+)}[\phi^{(s,k)}] \ \ \mbox{for}  \ 
   \Box\phi^{(s,k)} = 0 \ \} \ &\subset \ \cH_{\rm phys} \cap \cC^s  \nn\\
    \{c_-\tilde\cA^{(-)}[\phi^{(s,k)}] + \tilde\cA^{(n)}[\phi^{(s,k)}] \ \ \mbox{for}  \ 
   \Box\phi^{(s,k)} = 0 \ \} \ &\subset \ \cH_{\rm phys} \cap \cC^s  
   \label{Hphys-Cs-generic}
\end{align}
This completes the list of physical modes.

To summarize, 
the model contains generically 2 physical modes parametrized by $\phi^{(s)} \in\cC^s$ with $\Box \phi^{(s)} = 0$ for each spin $s\geq 2$,
up to the exceptional cases discussed above. These are ''would-be massive`` modes, i.e.
they contain the $2s+1$ dof of massive spin $s$ multiplets with vanishing mass parameter, and  decompose
further into a series of irreducible massless spin $s$ modes in radiation gauge for $k\leq s$
as described above. 
All modes transform and mix under a higher-spin gauge invariance.
It is hence plausible that some of these modes  become massive
in the interacting theory, but this remains to be clarified.

\subsection{Metric fluctuation modes}
\label{sec:graviton}

Now we discuss how  metric fluctuations arise from the above  
modes.
The effective metric for functions of $\cM^{3,1}$ on a perturbed background $Y = T + \cA$ can be extracted from the kinetic 
term  in \eqref{scalar-action-metric}, which defines the 
bi-derivation
\begin{align}
\begin{aligned}
 \g:\quad \cC\times \cC  \ &\to  \quad \cC  \\
  (\phi,\phi') &\mapsto \{Y^\a,\phi\}\{Y_\a,\phi'\} \ .
  \label{metric-full}
  \end{aligned}
\end{align}
Specializing to $\phi=x^\mu, \phi' = x^\nu$ we obtain  the coordinate form
\begin{align}
\g^{\mu\nu} &=  \obar\g^{\mu\nu} + \d_\cA \g^{\mu\nu} + [\{\cA^\a,x^\mu\}\{\cA_\a,x^\nu\}]_0
%  =  \obar\g^{\mu\nu} + \sinh(\eta)h^{\mu\nu} + \sinh^2(\eta)h^{\mu\a}\eta_{\a\b}h^{\nu\b}
\label{gamma-nonlinear}
\end{align}
 whose linearized contribution in $\cA$ is given by
\begin{align}
\begin{aligned}
 \d_\cA \g^{\mu\nu} &= \sinh(\eta)   h^{\mu\nu}[\cA], \qquad 
   h^{\mu\nu}[\cA] := \{\cA^\mu,x^\nu\}_0 + (\mu \leftrightarrow \nu)  \ .
  \label{gravitons-H1}
  \end{aligned}
\end{align}
and $h[\cA] = 2\{\cA^\mu,x_\mu\}_0$.
The projection on $\cC^0$ ensures that this is the metric for functions on $\cM^{3,1}$.
%We will focus on the linearized contribution  in the following.
Clearly only $\cA \in \cC^1$ can contribute, which we assume henceforth.
% To evaluate this explicitly,
% it is convenient to consider the following rescaled graviton mode: 
% \begin{align}
%   h^{\mu\nu}[\cA] &:=   \{\cA^\mu,x^\nu\}_0 + (\mu \leftrightarrow \nu)   \nn\\
%   % = -\cA^{\nu(-)}[\cA^\mu] + (\mu \leftrightarrow \nu), \nn\\
%  \qquad h[\cA] = 2\{\cA^\mu,x_\mu\}_0 \ .
%  \label{tilde-H-def}
% \end{align}
Including the conformal factor in \eq{eff-metric-G},
this leads to the effective metric fluctuation \cite{Steinacker:2019dii}
\begin{align}
 \d G^{\mu\nu} &= \b^2\ \tilde h^{\mu\nu} \ , 
 \label{eff-metric-fluct}
\end{align}
where 
\begin{align}
 \tilde h^{\mu\nu} = h^{\mu\nu} - \frac 12 \eta^{\mu\nu} h \ ,
 \qquad \b=\frac{1}{\sinh(\eta)} \ .
\end{align}
Let us discuss the mode content of $h^{\mu\nu}[\cA]$.
Recall that the 12 off-shell dof in $\cA_\mu = \cA_{\mu;\a} t^\a \in\cC^1$
are realized by $\cA^{(-)}[\phi^{(2)}]$,
$\cA^{(n)}[\phi^{(1)}], \cA^{(g)}[\phi^{(1)}]$ and 
$\cA^{(+)}[\phi^{(0)}]$. Hence the 10 dof of the 
most general off-shell metric fluctuations are provided by 
$\cA^{(-)}[\phi^{(2)}]$, 
$\cA^{(g)}[\phi^{(1)}]$, and the scalar modes $\cA^{(+)}[\phi^{(0)}]$
and $\cA^{(n)}[D^+\phi^{(0)}]$. 
%The physical metric fluctuations arise from $\cA^{(-)}[\phi^{(2)}]$.
According to the results of section \ref{sec:physical},
the  physical modes among these  are  the 5
would-be massive
spin 2 modes  $\cA^{(-)}[\phi^{(2)}]$, which decompose into the
massless graviton $\cA^{(-)}[\phi^{(2,0)}]$, 
one massless vector mode $\cA^{(-)}[\phi^{(2,1)}]$, and one scalar mode
$\cA^{(-)}[\phi^{(2,2)}]$.
The vector field can be extracted by
\begin{align}
 \{t_\mu,h^{\mu\nu}\} 
  &\stackrel{phys}{=} - \frac{2}{R} D^- \cA^\nu 
  \label{metric-div-general}
\end{align}
as shown in \cite{Steinacker:2019awe}, which
vanishes  for the $\cA^{(-)}[\phi^{(2,0)}]$ graviton mode.

\paragraph{Linearized Ricci tensor.}

To understand the significance of the metric modes, we consider the
linearized Ricci tensor 
\begin{align}
2 R_{(\rm lin)}^{\mu\nu}[G] 
 &= -\nabla^\a \nabla_\a \d G^{\mu \nu}  
 + \nabla^{\mu} \nabla_\r \d G^{\nu\r} + \nabla^{\nu} \nabla_\r \d  G^{\mu\r} 
  - \nabla^\mu\nabla^\nu \d G  
 \label{gmunu-lin}
\end{align}
for a metric fluctuation $\d G^{\mu\nu} = \b^2 \tilde h^{\mu\nu}$
around the background 
 $\obar G^{\mu\nu} = \b \eta^{\mu\nu}$.
 For simplicity, we  neglect contributions of the order of the cosmic background
 curvature. Then we can replace $\nabla$ by $\del$ in Cartesian coordinates, and \cite{Steinacker:2019dii}
\begin{align}
2R_{(\rm lin)}^{\mu\nu}[G] \
 &\stackrel{\eta\to\infty}{\approx} \ 
  \b^2\Big(-\del^\a\del_\a  \tilde h^{\mu\nu}
   + \del^\mu\del_\r  \tilde h^{\r\nu}
   + \del^\nu\del_\r  \tilde h^{\r\mu}
   - \del^\mu\del^\nu  \tilde h\Big) \nn\\
  &\stackrel{\rm on-shell}{\approx}  \ \b^2\Big(  \del^\mu\del_\r  h^{\r\nu} 
   + \del^\nu\del_\r h^{\r\mu} \ +  O(\frac{\del h^{\mu\nu}}{x_4})  \Big) \ .
   \label{Ricci-tensor-onshell}
\end{align}
Here the on-shell relation $(\cD^2-\frac{3}{R^2})\cA = 0$ is used as well as
 $\del \gg \frac 1{x_4}$ at late times $\eta\to\infty$,
 focusing on scales much shorter than the cosmic curvature scale.

\paragraph{Pure gauge modes.}
Consider first the metric fluctuation corresponding to the pure gauge fields $\cA^{(g)}[\phi]$, where 
$\phi \in \cC^{1}$  is a  spin 1 field. 
It is not hard to see that \eq{eff-metric-fluct} then takes the form
\begin{align}
  h^{\mu\nu}[\cA^{(g)}]  
  &= -\{t^\mu,\cA^{(-)\nu}\}
  -\{t^\nu,\cA^{(-)\mu}\}
  + 2  \eta^{\mu\nu} \{t^\a,\cA_\a^{(-)}\}\,,
 % \label{pure-gauge-metric} 
 \label{pure-gauge-H} 
\end{align}
where $\cA^{(-)}_\mu = \cA^{(-)}_\mu[\phi]$.
Then the pure gauge contribution to the effective metric \eq{eff-metric-fluct} is \cite{Steinacker:2019dii}
\begin{align}
 \d G^{\mu\nu}_{(g)}  &=  \nabla^\mu\xi^\nu + \nabla^\nu \xi^\mu, \qquad \xi^\mu = - \cA^{(-)\mu} 
 \label{puregauge-diffeo-rel}
\end{align}
Hence the pure gauge metric modes in the present framework can be identified with 
diffeomorphisms generated by $\xi = -\cA^{(-)} $. This provides a  non-trivial consistency check for the 
correct identification of the effective metric. 
These diffeomorphisms are essentially volume-preserving  due to
the constraint $\nabla_\a (\b^{3} \xi^\a) = 0$ \eq{gauge-diffeo-constraint},
% \begin{align}
%   \nabla_\a (\b^{3} \xi^\a) = 0 \ . 
%  \label{gauge-diffeo-constraint}
% \end{align}
leaving only 3 rather than 4 diffeomorphism d.o.f., unlike in GR.
This reflects the presence of a dynamical scalar metric degree of freedom, 
which we will discuss in more detail below.

\paragraph{The $\cA^{(-)}$ metric modes.}

Now consider the $\cA^{(-)}[\phi^{(s)}]$ modes. Among these,
only  the ones with spin $s=2$ can 
contribute to the metric, and these are precisely the physical degrees of freedom as shown above. 
The corresponding linearized metric fluctuation is \cite{Sperling:2019xar}
\begin{align}
  h^{\mu\nu}[\cA^{(-)}] 
 &= -2 \{x^\mu,\{x^\nu,\phi\}_-\}_- = -2 \{x^\nu,\{x^\mu,\phi\}_-\}_- \nn\\
 \eta_{\mu\nu}  h^{\mu\nu}
  &=  2 D_- D_-\phi  \, 
   = -\frac{1}{x_4^2} x_\mu x_\nu h^{\mu\nu}
    \label{Hmunu-tr}  
% % 
\end{align}
for $\phi \in \cC^2$. 
They satisfy the identity \cite{Steinacker:2019dii}
\begin{align}
 \del_\mu(\b h^{\mu\nu}) = 0 \ ,
  \label{A-h-nice-id}
\end{align}
which together with  \eqref{Ricci-tensor-onshell} implies that all these 
on-shell (would-be massive) spin 2 modes are Ricci-flat up to cosmic scales,
\begin{align}
 2R_{(\rm lin)}^{\mu\nu} \
 &= \  0 +  \  O(\frac{\del G^{\mu\nu}}{x_4}) \ .
   \label{Ricci-tensor-onshell-massive}
\end{align}
The  $\cA^{(-)}[\phi^{(2,0)}]$  modes  clearly reproduce the 
 two Ricci-flat graviton modes from GR.
 The modes arising from $\phi^{(2,1)}$ and  $\phi^{(2,2)}$ essentially complete a massive 
 spin 2 multiplet due to \eq{A-h-nice-id} and \eq{Hmunu-tr}, however the statement \eq{Ricci-tensor-onshell-massive} is largely empty, 
because these modes are generically  dominated by diffeos
which are trivially flat. 
Nevertheless, we will see that the {\em quasi-static} $\cA^{(-)}[\phi^{(2,2)}]$ solution
leads to a non-trivial Ricci-flat metric perturbation, which is 
nothing but the linearized Schwarzschild metric. 
This is consistent with the identity \eq{Hmunu-tr} for the effective metric \eq{eff-metric-fluct},
\begin{align}
 \eta_{\mu\nu}  \d G^{\mu\nu}
  &= \frac{2}{x_4^2} x_\mu x_\nu \d G^{\mu\nu} \ .
\end{align}

\subsection{The linearized Schwarzschild solution}

Now we work out the  metric perturbation arising from the 
physical $\cA^{(-)}[\phi^{(2,2)}]$ mode, where $\phi^{(2,2)} = D^+D\phi$
for  a scalar field $\phi\in\cC^0$.
This is part of the would-be massless spin 2 multiplet $\cA^{(-)}[\phi^{(2)}]$,
and the associated metric perturbation is guaranteed to be Ricci-flat (on-shell)
due to \eq{Ricci-tensor-onshell-massive}.
We will see that this includes  a quasi-static
Schwarzschild metric, as well as dynamical solutions which might be related to 
dark matter.
% The on-shell condition is $(\Box-\frac{4}{R^2})D^+D^+ \phi = 0$, which is 
% equivalent to 
% \begin{align}
% 0 = (\Box-\frac{4}{R^2})D^+D^+ \phi = D^+(\Box +\frac{4}{R^2} -\frac{4}{R^2})D^+ \phi
%  = D^+D^+ (\Box +\frac{2}{R^2})\phi
% \end{align}
We will use the on-shell condition 
$\Box \phi = - \frac{2}{R^2}\phi$,
and focus on the late-time limit $\eta\to\infty$.
% Then 
% \cite{}
% \begin{align}
%  \frac{5}{2r^2} h^{\mu\nu}[\cA^{(2)}[D^+D\phi]]
%   &= \frac 29 r^2 \Big(\eta^{\mu\nu}(2 + \t)(3 + \t^2+4\t)
%    +  \frac{\b^2}{R^2} x^\mu x^\nu (\t^2 -1)\nn\\
%    &\qquad   - (x^\nu\del^\mu + x^\mu\del^\nu)(\t^2+3\t+2) 
%     -  R^2\del^\nu  \del^\mu (\t+4)\Big)(\t+2)\phi \ .
% \end{align}
Then the trace-reversed metric fluctuation $\tilde h^{\mu\nu}$ is 
found to be \cite{Steinacker:2019dii}
\begin{align}
 \tilde h^{\mu\nu} 
 %&= h^{\mu\nu} - \frac 12 h \eta^{\mu\nu}  \nn\\
%   &= \frac{4r^4}{45} (\t+2)(\t+1)
%    \Big(\big((2 + \t)(\t + 3) - \frac 12 (2\t+5)(\t+5)  \big) \eta^{\mu\nu}  \nn\\
%    &\quad 
%     + \frac{\b^2}{R^2} x^\mu x^\nu (\t -1)
%     - (x^\nu\del^\mu + x^\mu\del^\nu)(\t+2)  \Big)\phi
%     -\frac{4r^4}{45}  R^2\del^\nu  \del^\mu(\t+4) (\t+2)  \phi  \nn\\
  &= \frac{4r^4}{45}
   \Big(-\frac 12 (5\t + 13) \eta^{\mu\nu} 
    + \frac{\b^2}{R^2} x^\mu x^\nu (\t -1)
    - (x^\nu\del^\mu + x^\mu\del^\nu) (\t+2) \Big) (\t+1)(\t+2)\phi \nn\\
 &\quad   -\frac{4r^4}{45}  R^2\del^\nu  \del^\mu   (\t+4)(\t+2)\phi \ .
    \label{tilde-h-nogauge}
\end{align}
Observe that for $\t\neq -2$,  the  term $(x^\nu\del^\mu + x^\mu\del^\nu)\phi$ is dominant 
at late times, since $x^0\sim R\cosh(\eta)$.
This can be removed using a suitable 
 diffeomorphism contribution 
$\nabla^\mu\xi^\nu + \nabla^\nu\xi^\mu$,  with the result
\begin{align}
 \tilde h^{\mu\nu} 
%   &\sim \frac{4r^4}{45} 
%    \Big(-\frac 12 (5\t + 13) \eta^{\mu\nu} 
%     + \frac{\b^2}{R^2} x^\mu x^\nu (\t -1)
%     +  \big(3\eta^{\mu\nu} + 2x^\nu x^\mu \frac{\b^2}{R^2}\big) (\t+2)\Big)(\t+1)(\t+2)\phi \nn\\
%    &\quad  -\frac{4r^4}{45}   R^2\b^2\del^\nu  \del^\mu (\t+4)(\t+2)\phi \nn\\
 %%
 &\sim \frac{4r^4}{45} 
   \Big( \frac 12(\t - 1)\eta^{\mu\nu}  
    + 3\frac{\b^2}{R^2} x^\mu x^\nu (\t +1)\Big)(\t+1)(\t+2)\phi 
    \label{eff-metric-after-diffeo}
\end{align}
for large $\eta$. Then 
\begin{align}
 \tilde h_{\mu\nu}\, d x^\mu d x^\nu 
 \ \ &=  \ \frac{2r^4R^2}{45} 
   \sinh^2(\eta) \big(d\eta^2(5\t +7) + d\Sigma^2(\t - 1)\big)(\t+1)(\t+2)\phi \nn\\
 &\stackrel{\t\to-2}{=} 
 %\ \frac{2r^4R^2}{15} \sinh^2(\eta)(\t+2)\phi\, \big(d\eta^2 + d\Sigma^2\big) \nn\\
 % &=
 - 4 \phi' (d t^2 + a(t)^2 d\Sigma^2)
\end{align}
using the explicit form \eqref{eff-metric-FRW} of the scale parameter $a(t)$ for large $\eta$,
where
\begin{align}
 \phi' = -\frac{r^4}{30} \b (\t+2)\phi \ .
 \label{phi-prime-new}
\end{align}
For $\t\neq -2$  
this metric is not Ricci-flat, which seems inconsistent with \eq{Ricci-tensor-onshell-massive}.
However then the diffeo contribution in \eqref{tilde-h-nogauge}
is very large at late times, invalidating the linearized approximation. 
Therefore we restrict ourselves to the ''quasi-static`` case $\t\approx -2$.
Then the full perturbed metric can be written 
in the form 
\begin{align}
 \label{perturbed-metric-cosm-expl}  
\boxed{\
\begin{aligned}
 ds^2 = (G_{\mu\nu} - \d G_{\mu\nu})  d x^\mu d x^\nu 
%   &= (\sinh(\eta)\eta_{\mu\nu} 
%     - \tilde h_{\mu\nu})\, d x^\mu d x^\nu \\
  &= - d t^2 + a(t)^2 d\Sigma^2 
      \ + 4 \phi' (d t^2 + a(t)^2 d\Sigma^2) \ .
 \end{aligned}
 \ }
\end{align}
The on-shell condition reduces to
$\Delta^{(3)}\phi = 0$ for $\t=-2$, 
and in the spherically symmetric case the Newton potential 
on a $k=-1$ geometry is recovered, more precisely
\begin{align}
 \phi \ &= \  \frac{e^{- \chi}}{\sinh(\chi)} \frac{1}{\cosh^2(\eta)}
 \ \sim \ \frac{1}{\r} e^{-\chi -2\eta} \ , \qquad \r = \sinh(\chi) \ .
 \label{phi-tis-2}
\end{align}
Strictly speaking we should  use $\phi'$ rather than $\phi$ in the $(\t+2)\phi = 0$ case; then
the quasi-static condition becomes $(\t+3+\b^2)\phi' = 0$ and the on-shell condition  is
$(\Delta^{(3)} - 4\b^4)\phi' = 0$, which reduces again to $\Delta^{(3)}\phi' = 0$
in the large $\eta$ limit. This gives
\begin{align}
  \phi' &\sim  \frac{1}{\r} e^{-\chi-3\eta} \ \sim  \frac{e^{-\chi}}{\r} \frac{1}{a(t)^2}
  \label{phiprime-solution}
\end{align}
 for large $\eta$, 
 recalling $a(t) \sim e^{-\frac 32 \eta}$ \eqref{a-eta}.
This metric is very close to the Vittie solution \cite{mcvittie1933mass} for the Schwarzschild metric 
for a point mass $M$ in a 
FRW spacetime, whose linearization for $k=-1$ is
\begin{align}
 ds^2 
   %&= -\Big(\frac{1-\mu}{1+\mu}\Big)^2 dt^2 + (1+\mu)^4 a(t)^2 d\Sigma^2  \nn\\
   &= -dt^2 + a(t)^2 d\Sigma^2  \ + \  4\mu(dt^2 + a(t)^2 d\Sigma^2)  \ + O(\mu^2) \ .
   \label{vittie}
 \end{align}  
Here 
\begin{align}  
  \mu = \mu(t,\chi) &= \frac{M}{2 \r} \frac{1}{a(t)} 
  \label{mu-vittie}
\end{align}
is the mass parameter, which is not constant but decays during the cosmic expansion; 
this is as it should be, because local gravitational systems do not participate in the 
expansion of the universe.
Comparing with  \eqref{phiprime-solution} we have
\begin{align}
  \phi' &\sim \  \mu(t,\chi)\, \frac{e^{-\chi}}{a(t)}  \ .
\end{align}
Since $\mu$ \eqref{mu-vittie}
looks like  a constant mass for a comoving observer  \cite{mcvittie1933mass}, 
the effective mass parameter in our solution effectively decreases like 
$a(t)^{-1}$ during the cosmic evolution.
This suggests a time-dependent
Newton constant, however this is premature since the 
coupling to matter is not fully understood, and quantum effects may modify the result.
Also, while both metrics have the characteristic $\frac 1\r$ dependence of the 
Newton potential, the present solution has an extra $e^{-\chi}$ factor, 
which reduces its range at cosmic  scales.
Both effects are irrelevant at solar system scales, 
but they will be important for cosmological considerations,  
reducing the gravitational attraction at long scales.

It is interesting that the scalar on-shell modes provide
a Ricci-flat metric perturbation only for the  quasi-static case 
$\t=-2$. Of course
it should be expected that a dynamical scalar metric mode, which does not exist in GR, is
not Ricci-flat in general. From a GR point of view, such non-Ricci-flat perturbations would
be interpreted as dark matter. One can argue \cite{Steinacker:2019dii} that they should be 
more important at very long wavelengths, however a more detailed examination 
at the non-linear level is needed to clarify the significance of the extra metric modes.

Finally, note that only vacuum solutions were considered so far. While the metric fluctuations couple to matter as usual, it is not evident
that the standard inhomogeneous Einstein equations arise; in fact some higher-derivative contributions of the type 
$(1+\Box_H)T^{\mu\nu}$  are expected at the classical level, cf. section  6.3 in \cite{Sperling:2019xar}.
On the other hand, quantum effects are bound to induce an Einstein-Hilbert-like term 
in the effective action \cite{Sakharov:1967pk}, and 
due to the (partial) local Lorentz invariance it is plausible that the resulting gravity theory
will be reasonably close to Einstein gravity. 
However, further work is needed to corroborate that claim.

\subsection{Towards quantization}
\label{sec:quantization}

We briefly comment on the quantization of the model via the matrix integral \eq{path-integral}.
As pointed out before, the maximal supersymmetry of the IKKT model leads to important cancellations 
in the loops, and the resulting gauge theory is very similar to $\cN=4$ SYM with  higher-spin-type gauge invariance.
It is thus reasonable to expect that the model is well-defined at the quantum level.
Nevertheless there are  issues to be considered. 
First, the $\cM^{3,1}$ background under consideration is non-compact
and described by matrices which are infinite-dimensional.
%Then the discussion around \eq{path-integral} is no longer rigorous.
Furthermore, the explicit mass term in \eq{MM-Mink} amounts to a soft SUSY 
breaking term, and will  reduce the degree of UV cancellations.
There is in fact a background-independent (albeit implicit) formula
for the one-loop effective action \cite{Ishibashi:1996xs,Chepelev:1997av,Blaschke:2011qu}
\begin{align}
\Gamma_{\!\textrm{1loop}}[X]\! 
%&= \frac 12 \Tr \Big(\log(\Box +\frac{\mu^2}2 - M^{(\cA)}_{ab}[ \Theta^{ab},.])
%-\frac 12 \log(\Box - M^{(\psi)}_{ab}[ \Theta^{ab},.])
%- 2 \log (\Box)\Big)   \nn\\
%  &= \frac 12 \Tr \Bigg(\sum_{n>0} \frac{1}n \Big((\Box^{-1}\big(-M^{(\cA)}_{ab}[ \Theta^{ab},.] 
%     + \frac{1}{2}\mu^2)\big)^n 
%   \, -\frac 12 (-\Box^{-1}M^{(\psi)}_{ab}[ \Theta^{ab},.])^n \Big)  \Bigg) \nn\\
  &= \frac 12 \Tr \Bigg(\!\! \frac 14 \Box^{-1}(M^{(\cA)}_{ab} [ \Theta^{ab},.] )^4 
  -\frac 18 (\Box^{-1}M^{(\psi)}_{ab} [ \Theta^{ab},.])^4 \,\, +  \cO(\Box^{-1}[ \Theta^{ab},.])^5 \! \Bigg) \nn\\
  &\quad + \frac 12  m^2 \Tr \Box^{-1} + O(m^4)
\label{Gamma-IKKT}
\end{align}
with  $a,b=0,...,9$. Here 
$M_{ab}^{(\psi)}$  and $M_{ab}^{(\cA)}$ are the $\mso(9,1)$ generators acting on the spinor and vector
representation, respectively.
This is basically the standard $\det=\exp\Tr\log$ formula, taking into account 
 cancellations due to SUSY.
Since the formula is background-independent, it provides the full one-loop 
effective action  by including fluctuations $\cA$ to the background \eq{covar-coords}.
The trace can be evaluated efficiently using the formalism of string states, 
cf. \cite{Steinacker:2016vgf,Steinacker:2016nsc}.
In the absence of the mass $m^2$, this is manifestly finite on $\cM^{3,1}$, as on any 
4-dimensional background\footnote{The internal $S^2$ fiber does not cause any complications 
because it is compact, and in fact $S^2_N$ admits only finitely many harmonics. 
Therefore the background is effectively 4-dimensional in the  UV.}.
However the last term leads to a divergence for $m^2\neq 0$.
Although this is just an irrelevant constant on the unperturbed background,
the cancellations are less effective, and
it would be better to find a finite-dimensional version of the  solution
which allows to scale $m^2$ with $N$ 
to obtain a well-defined large $N$ limit. This is also required to 
make contact with numerical simulations.
For a possible ansatz see  \cite{Steinacker:2017bhb}, but there may be better ones.
This is one of the open problems to be addressed in future work.

\subsection{Further literature}
\label{sec:lit}

We provide a selection of related literature which may be useful to understand the present 
framework and its broader context. Fuzzy spaces were introduced in
\cite{hoppe1982QuaTheMasRelSurTwoBouStaPro,Madore:1991bw,Grosse:1995ar}, 
and useful discussions from a field theoretical point of view can be found e.g. in 
\cite{Balachandran:2005ew,Ydri:2016dmy,Steinacker:2011ix,Balachandran:2001dd}. 
More mathematical details on quantized symplectic spaces can be found  in
\cite{Ali:2004ft,ma2008toeplitz,Bordemann:1993zv}, and for quantized coadjoint orbits 
see e.g. \cite{Hawkins:1997gj} or section 4.2 in \cite{Pawelczyk:2002kd}.
Coherent states on fuzzy spaces are very useful  
 \cite{Grosse:1993uq,Perelomov:1986tf,Steinacker:2016nsc,Ishiki:2015saa}, 
and provide the basis for a visualization tool   \cite{lukas_schneiderbauer_2016_45045}.
Fuzzy spaces as solutions of Yang-Mills matrix models 
and the associated  NC gauge theory have been studied e.g. in 
\cite{Alekseev:2000fd,Iso:2001mg,Kimura:2002nq,Steinacker:2003sd,Kimura:2003ab,Azuma:2004qe,
Behr:2005wp,Grosse:2004wm,Chaney:2015ktw,CastroVillarreal:2005uu},
and for NC field theory more generally see 
\cite{Douglas:2001ba,Szabo:2001kg} and references therein.
The role of fuzzy spaces as D-branes in string theory is discussed in 
\cite{Myers:1999ps,Felder:1999ka,Alekseev:2000fd,Fredenhagen:2000ei,Pawelczyk:2002kd}, and
in the context of matrix quantum mechanics e.g. in 
\cite{Banks:1996vh,Nair:1998bp,Sochichiu:2005ex,Hoppe:2002km,Kabat:1997sa,Castelino:1997rv}.
A relation of NC gauge theory and (emergent) gravity has long been suspected, 
and the effective metric in matrix models and its dynamical aspects
are discussed in \cite{Steinacker:2008ri,Steinacker:2010rh,Yang:2004vd,Rivelles:2002ez}.
Fuzzy extra dimensions are based on similar structures and can provide a 
relation with particle physics, see e.g. 
\cite{Chatzistavrakidis:2011gs,Aschieri:2006uw,Sperling:2018hys,Aoki:2014cya}. 

Covariant higher-dimensional fuzzy spaces were studied e.g. in 
\cite{Grosse:1996mz,Heckman:2014xha,Ramgoolam:2001zx,Zhang:2001xs,deMedeiros:2004wb,Dolan:2003th,Kimura:2002nq,Steinacker:2015dra,
Steinacker:2016vgf,Sperling:2017gmy,Manolakos:2019fle,Fiore:2017ude}, 
which are similar in spirit to Snyder space 
\cite{Snyder:1946qz,Yang:1947ud}, see also \cite{Hanada:2005vr} for a somewhat related ansatz.
In particular, the relation  of fuzzy $S^4$ to fuzzy $\C P^2$ was 
pointed out in \cite{Ho:2001as,Medina:2002pc,Karabali:2003bt},
which is analogous to the bundle structure discussed in section \ref{sec:fuzzyH4}. 
%The idea of using Lie algebras and their projections  to find solutions 
%of the matrix eom such as \eq{eom} is very useful, see e.g.  \cite{Kim:2012mw,Hoppe:2002km,Chatzistavrakidis:2011su,Steinacker:2014lma}.
For numerical investigations we refer to \cite{Anagnostopoulos:2013xga,Kim:2011cr,Nishimura:2019qal,Tekel:2015zga} and references therein.

\section{Conclusion and outlook}

The short summary of this review article is that an explicit model for 3+1-dimensional quantum space-time 
has been established,
which is a solution of the mass-deformed IKKT-type matrix model, and  leads to a 
consistent higher-spin gauge theory without ghosts.
While local Lorentz invariance is only partially manifest, it appears to be respected effectively.
This theory includes a dynamical metric leading to an emergent gravity model, 
which includes the standard propagating massless spin 2 gravitons,
as well as the linearized Schwarzschild solution. 
However the metric also includes extra physical dof which can be viewed as arising from a would-be massive spin 2 mode.
While the full dynamics of the emergent gravity and  its dependence on matter is yet to be understood, 
the results so far make it plausible that it will be reasonably close to Einstein gravity, 
at least upon taking into account quantum effects.

The matrix model framework thus provides a unified treatment of space-time and field theory, 
and the present solution provides arguably the most satisfactory background for the model so far.
However its study is just at the beginning, and many things need to be worked 
out and clarified.

An important question which should be addressed is the following: what are the observable signature of this scenario,
and are there any clear-cut signals which would distinguish it from other approaches?
Of course this question can only be answered reliably once the resulting gravitational physics
is worked out, which is the most urgent open problem. 
This is not  trivial because it may be 
necessary to include quantum effects 
(in the sense of induced gravity a la Sakharov \cite{Sakharov:1967pk}), 
and/or to find a way to break the higher-spin gauge invariance beyond spin 2. 
Another, perhaps more clear-cut open problem is to find an exact non-linear analog of the (classical) Schwarzschild solution
extending the present linearized solution. This would  give important information about the 
formation of horizons in the present framework, and hints about the resolution of singularities.
Similarly, the presence of a classical Big Bounce solution is very intriguing, and could certainly be explored 
further in the present stage. Modifications of the time evolution should be possible by slightly relaxing the 
Lie algebra structure, and even compactified solutions are conceivable. 
In particular, it would be  desirable to find a FLRW solution with $k=0$.
Finally, closer links with Vasilievs higher spin gravity as 
well string theory realizations of the present brane solutions should be studied.
These are only some of the possible directions for further work.

\subsection*{Acknowledgments}

This review is based on talks given at several meetings including at the EISA Corfu, 
IHES Bures-sur-Yvette, NCTS Hsinchu, and the ESI Vienna.
I would like to thank the organizers of these meetings for providing the opportunity to meet and discuss with many 
people including C-S Chu, T. Damour, S. Fredenhagen,  J. Hoppe, C. Iazeolla, J. Karczmarek, H. Kawai,  J. Nishimura,
E. Skvortsov, and A. Tsuchiya.
Part of the work underlying this review was done in collaboration with M. Sperling, 
which is gratefully acknowledged. 
This work was supported by the Austrian Science Fund (FWF) grant P32086-N27.

\bibliographystyle{JHEP}
\bibliography{papers}

\end{document}